\documentclass[12pt,onecolumn,floatfix,superscriptaddress,tightenlines,nofootinbib]{revtex4-1}
\pdfoutput=1
\usepackage[colorlinks=true,citecolor=blue,linkcolor=blue,breaklinks=true]{hyperref}
\usepackage{amsmath,amssymb}
\usepackage{epsfig} 
\usepackage{graphicx}        
\usepackage{url}
\usepackage{color}
\usepackage{multirow}
\usepackage{placeins}
\usepackage[dvipsnames]{xcolor}
\usepackage{braket}
\usepackage{float}
\usepackage{slashed}
\hypersetup{
  colorlinks=true,
  linkcolor=red,
  filecolor=magenta,   
  urlcolor=blue,
  citecolor=blue,
} 
\providecommand{\xlink}[1]
 {\href{http://arxiv.org/abs/#1}{arXiv:#1}}
 
\allowdisplaybreaks

\bibpunct{[}{]}{,}{n}{}{,}

\def\beq{\begin{equation}}
\def\eeq{\end{equation}}
\def\bea{\begin{eqnarray}}
\def\eea{\end{eqnarray}}
\makeatletter
\@addtoreset{equation}{section}
\renewcommand{\theequation}{\thesection.\arabic{equation}}

\def\thesection{\Roman{section}}

\begin{document}

\title{Flavour effects in  gravitational leptogenesis }
\author{Rome Samanta}
\email{romesamanta@gmail.com}
\affiliation{Physics and Astronomy, University of Southampton, Southampton, SO17 1BJ, U.K.}
\author{Satyabrata Datta}
\email{satyabrata.datta@saha.ac.in}
\affiliation{Saha Institute of Nuclear Physics, HBNI, 1/AF Bidhannagar,
Kolkata 700064, India}


\begin{abstract} 
Within the Type-I seesaw mechanism, quantum effects of the right-handed (RH) neutrinos in the gravitational background lead to an asymmetric propagation of lepton and anti-leptons which allows a Ricci scalar and  neutrino Dirac-Yukawa coupling dependent chemical potential and therefore a lepton asymmetry in equilibrium. At high temperature, lepton number violating scattering processes try to maintain a dynamically generated lepton asymmetry in equilibrium. However, when the temperature drops down, the interactions become weaker, and the asymmetry freezes out. The frozen out asymmetry can act as a pre-existing asymmetry prior to the standard Fukugita-Yanagida leptogenesis phase ($T_i\sim M_i$, where $M_i$ is the mass of $i$th RH neutrino).   It is then natural to consider the viability of gravitational leptogenesis for a given RH mass spectrum which is not consistent with successful leptogenesis from decays. Primary threat to this gravity-induced lepton asymmetry to be able to successfully reproduce the observed baryon-to-photon ratio is the lepton number violating washout processes at $T_i\sim M_i$. In a minimal seesaw set up with two RH neutrinos, these washout processes are strong enough to erase a pre-existing asymmetry of significant magnitude. We show that when effects of flavour on the washout processes are taken into account, the mechanism opens up the possibility of successful leptogenesis (gravitational) for a mass spectrum $M_2\gg 10^9 {\rm GeV}\gg M_1$ with $M_1 \gtrsim 6.3 \times 10^6$ GeV. We then briefly discuss how, in general, the mechanism leaves its imprints on the low energy  CP phases and absolute light neutrino mass scale.

\end{abstract}

\maketitle

\section{Introduction}
The dominance of matter over antimatter remains one of the outstanding questions in particle physics and cosmology. A simple and widely studied approach to this end is to create a lepton asymmetry and process it to the baryon asymmetry through $B-L$ conserving sphalerons\cite{sph1,sph2}. The seesaw mechanism\cite{sw1,sw2,sw3} which gives rise to the observed $\sim \rm eV$ scale\cite{pdg} light neutrino masses also facilitates lepton number violating processes in the early universe. Within this mechanism, lepton number and CP-violating decays of heavy right-handed (RH) Standard Model (SM) singlets when accompanied with out of equilibrium condition\cite{Sakharov:1967dj}, create a lepton asymmetry (leptogenesis)\cite{fuku,pilaf,bari,nir,rio} which is then converted to baryon asymmetry (baryogenesis/ matter-antimatter asymmetry) by sphaleron transition. Barring the SM gauge symmetry, in a most general scenario where the seesaw model is not subjected to any other symmetry (e.g., flavour symmetry\cite{fasy1,fasy2,fasy3,fasy4}), it is natural to assume that the heavy RH states are hierarchical. It is then easy to show that the minimum RH mass scale pertaining to a successful leptogenesis is $M_1\sim 10^9$ GeV (Davidson-Ibarra (DI) bound\cite{ibarra}) which is beyond the reach of the collider experiments. Obtaining testable predictions from leptogenesis thus requires either a lowering of the RH mass scale and going beyond the hierarchical limit or reduction in the number of free model parameters so that it can be tested indirectly in low energy neutrino experiments. To this end, whilst for a direct test, mechanisms such as leptogenesis from RH neutrino oscillation\cite{Ars}, a recently proposed mechanism of leptogenesis from Higgs decays\cite{Ham} and resonant leptogenesis due to strongly quasi-degenerate heavy neutrinos\cite{pilaf,dev1,dev2} are quite promising, for the latter, leptogenesis in grand unified theories like SO(10)\cite{so1,so2,so3,so4,so5} is worthwhile to give an emphasis on.\\ 

\begin{figure}
\includegraphics[scale=.5]{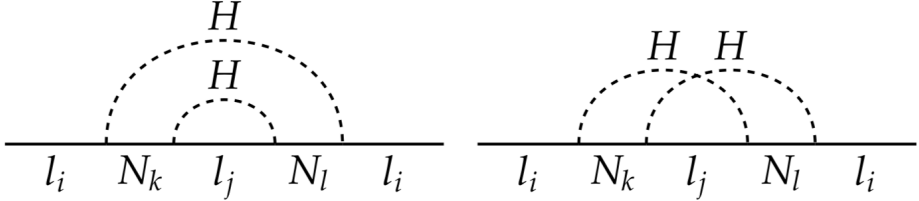}
\caption{Two-loop diagrams in seesaw model that generate the $\partial_\mu R j^\mu/M^2$  operator, e.g., see Ref.\cite{rg9}.}\label{fig1}
\end{figure}

A different perspective in the leptogenesis scenario has also been introduced by considering the interplay of particle physics and gravity where the lepton asymmetry is not produced by the decays or oscillation of particles rather the asymmetry is sourced by gravitational interactions. For example, lepton asymmetry sourced by chiral Gravitational Waves (GW)\cite{gravlep1,gravlep2,Caldwell:2017chz,gravlep3,Papageorgiou:2017yup,Kamada,gravlep4} and by the interaction of lepton or baryon current with background gravity through a C and CP- violating operator $\partial_\mu R j^\mu/M^2$, where $R$ is the Ricci scalar\cite{rg1,rg2,tra,Li:2004hh,Feng:2004mq,Lambiase:2006md,rg3,rg4,Lambiase:2013haa,rg5,rg6}. The lepton asymmetry from  GW is a  consequence of a chiral imbalance in the SM model which sources the asymmetry in the form of left- handed neutrinos, thus not easy to be realised in seesaw models\cite{gravlep3}. On the other hand, the operator  $\partial_\mu R j^\mu/M^2$ can be generated in seesaw models at two-loop level\cite{rg7,rg8,rg9,rg10} (cf. Fig.\ref{fig1}) causing a chemical potential and hence a net lepton asymmetry in equilibrium proportional to the time derivative of $R$. The physical reason for the production of lepton asymmetry in this scenario could be attributed to the fact that C and CP violating operators when couple to the curvature at the quantum level, lead to asymmetric propagation (create a difference in lepton and anti-lepton self-energy) of matter and anti-matter--a phenomenon which is forbidden in flat space by translation and CPT invariance\cite{rg7}. Starting from a minimally coupled Type-I seesaw Lagrangian
\bea
-\mathcal{L}^{\rm seesaw}=\sqrt{-g}\left[\bar{N}_{Ri} \slashed{D}N_{Ri}+ f_{\alpha i}\bar{\ell}_{L\alpha}\tilde{H} N_{Ri}
+\frac{1}{2}\bar{N}_{Ri}^C(M_R)_{ij} \delta _{ij}N_{Rj} 
+ {\rm h.c.}\right]\,, \label{seesawlag0}
\eea
where $\sqrt{-g}$ is the square root of the metric determinant, $f$ is the neutrino Dirac-Yukawa coupling, $l_{L\alpha}=\begin{pmatrix}\nu_{L\alpha} & e_{L\alpha}\end{pmatrix}^T$ is the SM lepton doublet of flavour $\alpha$, $\tilde{H}=i\sigma^2 H^*$ with $H= \begin{pmatrix}H^+&H^0
\end{pmatrix}^T $ being the Higgs doublet and $M_R={\rm diag}\hspace{.5mm} (M_1,M_2,M_3)$, $M_{1,2,3}>0$, the generated equilibrium asymmetry at a temperature $T$ is given by\cite{rg9}
\bea
N_{B-L}^{eq}=\frac{\pi^2\dot{R}}{36 (4\pi )^4}\sum_{j>i}\frac{{\rm Im}\left[k_{ij}^2\right]}{\zeta (3)T M_i M_j}\left(\frac{M_j^2}{M_i^2}\right)^p{\rm ln }\left(\frac{M_j^2}{M_i^2}\right),\label{gl1}
\eea

where $k_{ij}=(f^\dagger f)_{ij}$ and $p=0,1$\cite{rg8,rg9}. A  dynamically generated asymmetry then freezes out once the relevant interactions (non-resonant relativistic $N_i$-exchange or $\Delta L=2$ processes) that try to maintain the asymmetry in equilibrium become weaker. Although for $p=1$, the equilibrium asymmetry and hence the frozen out asymmetry ($N_{B-L}^{G0}$) get enhanced hierarchically\cite{rg9}, in a generic seesaw model it is not trivial to realise a pure RH neutrino induced gravitational leptogenesis (or following Ref.\cite{rg10},  Radiatively-induced gravitational leptogenesis (RIGL)).  The reasons being, firstly, the gravitationally produced asymmetry competes with the asymmetry produced by RH neutrino decays, i.e., one has to distinguish the parameter space of each of the cases. On the other hand, even if by a suitable choice of RH mass spectrum one underestimates the contribution from RH neutrino decays towards successful leptogenesis, at the standard thermal leptogenesis phase ($T\sim M_i$), the lepton number violating washout processes which are always present in a seesaw model in general wash out any pre-existing asymmetry\cite{pre1,pre2}, here $N_{B-L}^{G0}$, exponentially. 

The latter one is a matter of concern in the present work. We stick to the minimal requirement of two RH neutrinos ($N_1$ and $N_2$)\cite{ms0,ms1,ms2,ms3} to generate light neutrino masses and choose the RH mass spectrum such that the decays from both the RH neutrinos ($N_{1,2}$-leptogenesis) do not suffice to reproduce the correct baryon asymmetry. Thus we are left only with the asymmetry generated by the gravitational interaction of RH neutrinos, which then faces washout by the lepton number violating $N_i$-interactions ($N_i$-inverse decays). The final asymmetry can be represented   mathematically as
\bea
N_{B-L}^{Gf}\sim N_{B-L}^{G0}\mathcal{D}([f^\dagger f ]_{ii}),\label{di}
\eea
where $\mathcal{D}$ encodes an exponential dilution of the produced asymmetry by the washout processes. Strength of $\mathcal{D}$ then dictates the fate of successful gravitational leptogenesis. If $N_{B-L}^{G0}$ is produced at a temperature $T_0 \gg M_i$ (here $i=1,2$), it then faces $`i$' number of washout at the scales $T_i \sim M_i$  by $N_i$-interactions\cite{lu1,lu2}. In an unflavoured scenario (cf. Eq.\ref{di}), these washout effects are strong enough in minimal seesaw model (e.g., for a normal light neutrino mass ordering, $\mathcal{D}_{\rm min} \sim e^{-10^3 m_2/{\rm eV}}$, $m_2$ being the lightest non-zero light neutrino mass and one has $m_3>m_2>m_1=0$.) to erase $N_{B-L}^{G0}$. However, when effects of fast charged lepton interactions, i.e., interactions of lepton doublets with the RH component of charged leptons -- popularly known as flavour effects in leptogenesis\cite{fl1,fl2,fl3,fl4,fl5,fl6,Samanta:2018efa}, on the washout processes are accounted for, one has to track the asymmetry in relevant flavours, and Eq.\ref{di} can be generalised to\footnote{Please note that, this mathematical form is naive and given only for the introduction purpose. In a realistic scenario, it requires more detailing which are discussed in relevant places.} 
\bea
N_{B-L}^{Gf}\sim \sum_\alpha N_{B-L}^{G0}\mathcal{D}(|f_{i\alpha}|^2).\label{di2}
\eea
 Given the current neutrino oscillation data, we show that the strength of $\mathcal{D}(|f_{i\alpha}|^2)$ can be reduced drastically (e.g., dominantly in the electron flavour for normal light neutrino mass ordering) and consequently, $N_{B-L}^{G0}$ does not face significant washout at $T_i \sim M_i$.  This opens up the possibility to obtain pure gravitational leptogenesis in minimal seesaw models. 
\begin{figure}
\includegraphics[scale=.5]{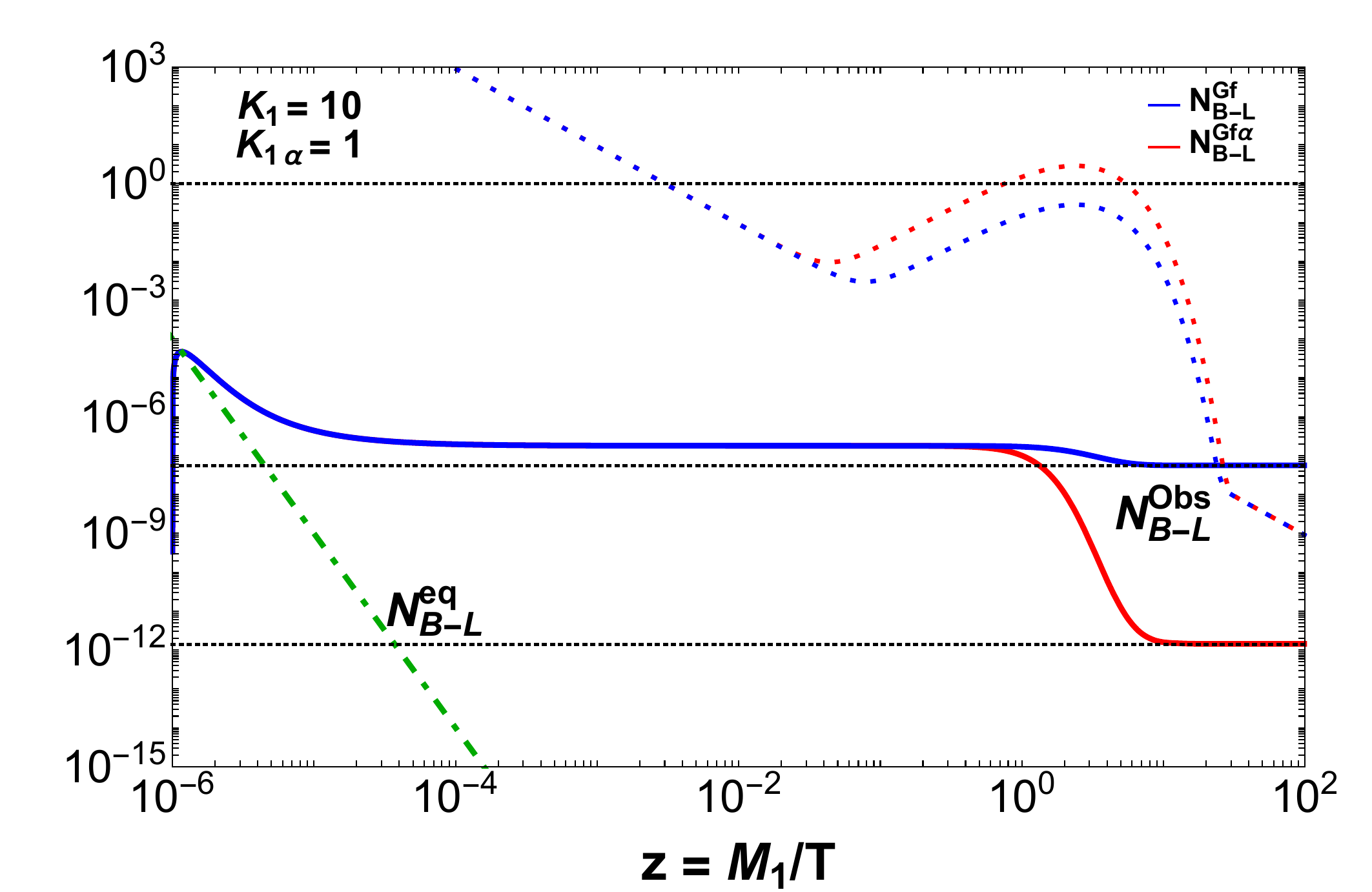}
\caption{Evolution of Gravity induced lepton asymmetry for a benchmark value of $M_1=10^7~ {\rm GeV}$. The solid blue (red) line is the flavoured (unflavoured) asymmetry. The dashed blue (red) line is the flavoured (unflavoured) washout rate. }\label{fig2}
\end{figure}
Specifically,  we consider two different  hierarchical spectrum of RH masses, a)  $10^9 {\rm GeV}\ll M_2\ll 10^{12} {\rm GeV}\lesssim T_0$ , $M_1\ll 10^9$  GeV, i.e., $M_2$ is in the two flavour regime and $M_1$ is in the three flavour regime (we shall explain flavour regimes in detail in Sec.\ref{s3}) b) $ 10^{12} {\rm GeV}\lesssim T_0\ll M_2$, $M_1\ll 10^9$  GeV, i.e., $M_2$ is in the unflavoured (one flavour) regime and $M_1$ is in the three flavour regime. For these spectrum of masses, it is well known that lepton asymmetry produced by RH neutrino decays is not adequate enough to be consistent with the observed baryon asymmetry, see e.g.,\cite{ibarra,2rh1,2rh2,2rh3}. However, as mentioned earlier, $N_{B-L}^{G0}$ which is produced gravitationally survives the washout effects owing to the fast charged lepton interactions which reduce the strength of the dilution factor $\mathcal{D}$.  After a detail quantitative study of flavour effects, we show that the spectrum b) with a normal light neutrino mass ordering (which is now favoured by neutrino oscillation data) facilitates successful gravitational leptogenesis and the lightest RH mass scale can be lowered to $\sim 6.3\times 10^6$ GeV. Thus overall, our results have a two-fold impact on the studies related to leptogenesis in seesaw models. Firstly, possibility of successful leptogenesis with the lightest  RH mass scale  below DI bound on $M_1$ ($M_1\gtrsim10^9$ GeV)\cite{ibarra}. Secondly, in minimal seesaw models, a new (non-standard) spectrum of RH masses emerge that reproduces correct baryon asymmetry.  Of course, the production mechanism of the lepton asymmetry is now different  -- the asymmetry does not originate from the RH neutrino decays, rather it originates due to the quantum effects of RH neutrinos in a gravitational environment and the key difference is, that unlike the traditional leptogenesis scenario, here the production and washout occur in different flavour regimes. In a nutshell,  entire discussion in this paper can be naively interpreted through Fig.\ref{fig2} (which will be more detailed in Sec.\ref{s3}). As one sees, after the departure from the equilibrium the flavoured (blue) asymmetry does not encounter significant washout and explains the observed baryon asymmetry ($N_{B-L}^{\rm Obs}$) whereas the unflavoured (red) faces a strong washout at e.g., $T_1\sim M_1$ and fails to reproduce $N_{B-L}^{\rm Obs}$. In fact, we will see later that for the flavoured case, the dilution factor  $D\sim e^{-K_{1\alpha}}$ and there exists a large parameter space with $K_{1\alpha}\ll 1$ so that practically there is no washout at $T_1\sim M_1$, however, in unflavoured case $D\sim e^{-K_{1}}$ and $K_1\gg 1$ thus the washout is strong. In either case, $K_{i(\alpha)}$ is called washout or decay parameter and is a function of $N_i$-Yukawa couplings.\\
 
The rest of the paper is organised as follows: In Sec.\ref{s2}, we discuss unflavoured leptogenesis for RH neutrino decays as well as for gravitational interaction. In Sec.\ref{s3}, we discuss flavoured leptogenesis scenario and show how in two RH neutrino seesaw model, a purely gravitational leptogenesis is realised. In Sec.\ref{s4}, we presented a detail numerical study and discussed the impact of RIGL mechanism on low energy neutrino observables. In Sec.\ref{s5} we summarise our results.
\section{One flavour leptogenesis in two RH neutrino seesaw model}\label{s2}
In this section we briefly discuss leptogenesis from RH neutrino decays in the presence of a pre-existing lepton asymmetry created by the quantum effects of RH neutrinos in gravitational background. In a  two RH neutrino seesaw model at a temperature $T_{B1}<M_1$ where the $N_1$ interactions go out of equilibrium, the final asymmetry can be written as
\bea
N_{B-L}^f=N_{B-L}^{Gf}+N_{B-L}^{Df}.\label{nblt}
\eea
The  number densities in Eq.\ref{nblt} are  normalised  to the co-moving number density of photons\cite{bari}. The first term is a contribution that originates due to the gravitational interactions of the RH neutrinos (after all the relevant washouts end) and the second term arises from the RH neutrino decays. Assuming the standard thermal history of the universe, the final baryon to photon ratio can be written as 
\bea
\eta_B = a_{\rm sph}\frac{N_{B-L}^f}{f_{\gamma}}\simeq  10^{-2}N_{B-L}^f,\label{theta}
\eea
where $f_{\gamma}$ is the  photon dilution factor  and $a_{\rm sph}\sim 1/3$ is the sphaleron conversion coefficient\cite{bari}. For a successful leptogenesis one has to compare Eq.\ref{theta} to the observed value $\eta_{\rm CMB}\sim (6.3\pm 0.3)\times 10^{-10}$\cite{planck}. 
First, we discuss the generation of the $B-L$ asymmetry $N_{B-L}^{Df}$ from RH neutrino decays. Starting  from the neutrino mass terms in the  seesaw Lagrangian in Eq.\ref{seesawlag0}
\bea
-\mathcal{L}_{mass}^{\nu,N}= \bar{\nu}_{L\alpha}(m_D)_{i\alpha}N_{Ri}
+\frac{1}{2}\bar{N}_{Ri}^C(M_R)_{ij} \delta _{ij}N_{Rj} 
+ {\rm h.c.}\,, \label{seesawlag}
\eea
where $m_D=fv$ with $v=174$ GeV being the vacuum expectation value of the SM Higgs, the effective light neutrino mass matrix can be obtained with the seesaw mechanism\cite{sw1} as
\bea
M_\nu = -m_DM_R^{-1}m_D^T\,. \label{seesaweq}
\eea
The mass matrix in Eq.\ref{seesaweq} can be diagonalised by a unitary matrix $U$ as
\bea
U^\dagger m_D M_R^{-1}m_D^T U^*=D_m,\label{see2}
\eea
where $D_m=-~{\rm diag}~(m_1,m_2,m_3)$ with $m_{1,2,3}$ being the physical light neutrino masses. We work in a basis where the RH neutrino mass matrix $M_R$ and  charged lepton mass matrix $m_\ell$  are diagonal. Thus, the neutrino mixing matrix $U$ can be written as 
\bea
U=P_\phi U_{PMNS}\equiv 
P_\phi \begin{pmatrix}
c_{1 2}c_{1 3} & s_{1 2}c_{1 3} & s_{1 3}e^{-i\delta }\\
-s_{1 2}c_{2 3}-c_{1 2}s_{2 3}s_{1 3} e^{i\delta }&  c_{1 2}c_{2 3}-s_{1 2}s_{1 3} s_{2 3} e^{i\delta} & c_{1 3}s_{2 3} \\
s_{1 2}s_{2 3}-c_{1 2}s_{1 3}c_{2 3}e^{i\delta} &-c_{1 2}s_{2 3}-s_{1 2}s_{1 3}c_{2 3}e^{i\delta} & c_{1 3}c_{2 3}
\end{pmatrix}P_M\,,\nonumber\\
\label{eu}
\eea
where $P_M={\rm diag}~(e^{i\alpha_M},~1,~e^{i\beta_M})$  is the Majorana phase matrix,  $P_\phi={\rm diag}~(e^{i\phi_1},~e^{i\phi_2},~e^{i\phi_3})$  is an unphysical diagonal phase matrix and $c_{ij}\equiv\cos\theta_{ij}$, $s_{ij}\equiv\sin\theta_{ij}$ with the mixing angles $\theta_{ij}=[0,\pi/2]$.  CP violation enters in Eq.\,\ref{eu} through the Dirac phase $\delta$ and the Majorana phases $\alpha_M$ and $\beta_M$. It is convenient to parametrise (which can be straightforwardly derived from Eq.\ref{see2}) the Dirac mass matrix as
\bea
m_D=U\sqrt{D_m}\Omega\sqrt{M_R},\label{orth}
\eea
where $\Omega$ is a $3\times 3$ complex orthogonal matrix. As an aside, in Table \ref{oscx}, let's present the latest fact file for the light neutrinos.
\begin{table}[H]
\begin{center}
\caption{Input values used in the analysis (inclusive of SK data)\cite{globalfit}} \label{t1}
\vspace{2mm}\label{oscx}
 \begin{tabular}{|c|c|c|c|c|c|}
\hline
\hline
${\rm Parameter}$&$\theta_{12}$&$\theta_{23}$ &$\theta_{13}$ &$ \Delta
m_{21}^2$&$|\Delta m_{31}^2|$\\
&$\rm degrees$&$\rm degrees$ &$\rm degrees$ &$ 10^{-5}\rm
(eV)^2$&$10^{-3} \rm (eV^2)$\\
\hline
$3\sigma\hspace{1mm}{\rm
ranges\hspace{1mm}(NO)\hspace{1mm}}$&$31.61-36.27$&$41.1-51.3$&$8.22-8.98$&
$6.79-8.01$&$2.44-2.62$\\
\hline
$3\sigma\hspace{1mm}{\rm
ranges\hspace{1mm}(IO)\hspace{1mm}}$&$31.61-36.27$&$41.4-51.3$&$8.26-9.02$&
$6.79-8.01$&$2.42-2.60$\\
\hline
${\rm Best\hspace{1mm}{\rm fit\hspace{1mm}}values\hspace{1mm}(NO)}$ &
$33.82$ & $48.6$ &  $8.60$ &$7.39$ & $2.53$\\
\hline
${\rm Best\hspace{1mm}{\rm
fit\hspace{1mm}}values\hspace{1mm}(IO)}$&$33.22$&$48.8$&$8.64$&$7.39$&$2.51$\\
\hline
\end{tabular}
\end{center}
\end{table}
Unlike the previously released data\cite{gfpre}, present best-fit value ($\sim 221^o$) for the Dirac CP violating phase ($\delta$) exhibits a shift towards its CP conserving value for the Normal mass Ordering (NO), though for the Inverted mass Ordering (IO), best-fit of $\delta$ is still close to its maximal value ($\sim 282^o$). The Majorana phases remain unconstrained and there is a preference of a Normal Ordering (NO) over an Inverted Ordering (IO).\\

Before the Electroweak Symmetry Breaking (EWSB), the RH neutrinos decays to lepton doublets and Higgs (cf. Eq.\ref{seesawlag0}). The produced lepton doublets $\ket{\ell_i}$  can be written as a coherent superposition of the corresponding flavour states $\ket{\ell_{\alpha}}$ as,
\bea
\ket{\ell_i}&=&\mathcal{A}_{i\alpha} \ket{\ell_\alpha} \hspace{1cm} (i=1,2; \alpha=e,\mu,\tau)\label{coh1}\\
\ket{\bar{\ell}_i}&=&\bar{\mathcal{A}}_{i\alpha} \ket{\bar{\ell}_\alpha} \hspace{1cm} (i=1,2; \alpha=e,\mu,\tau)\,,\label{coh2}
\eea
where the tree-level amplitudes are given by
 \bea
 \mathcal{A}_{i\alpha}^0 =\frac{m_{D_{i\alpha}}}{\sqrt{(m_D^\dagger m_D)_{ii}}}\hspace{1cm}{\rm and}\hspace{1cm}\bar{\mathcal{A}}_{i\alpha}^0 =\frac{m^*_{D_{i\alpha}}}{\sqrt{(m_D^\dagger m_D)_{ii}}}.\label{states}
 \eea
 The asymmetry produced by the CP-violating decays of the RH neutrinos is given by
\bea
N_{B-L}^{Df}=\frac{3}{4}\sum_{i}^2\varepsilon_{i}\kappa_{i}\,.
\label{fep}
\eea
where $\varepsilon_i$ is the CP asymmetry parameter a nonzero value of which is ensured by the complex phases in the matrices $U$ and $\Omega$. The efficiency factor
\bea
\kappa_{i} (z=M_1/T)=-\frac{4}{3}\int_{z_{\rm in}}^z \frac{dN_{N_i}}{dz^\prime}e^{-\sum_{j}\int_{z^\prime}^z W_{j}(z^{\prime\prime})dz^{\prime\prime}}dz^\prime\,,\label{effi}
\eea
contains the information of washout processes involving the inverse decays and lepton number violating scattering processes\cite{bari,fl5}. At a temperature $z_{B1}\sim M_1$ the $N_1$-washout processes go out of equilibrium and in the hierarchical limit $M_2\gg M_1$  the efficiency factors for thermal initial abundance of the RH neutrinos can be  computed as\cite{bari}
\bea
\kappa_1^\infty &=&\frac{2}{K_1 z_B (K_1)} \left(1-e^{-\frac{K_1 z_B(K_1)}{2}}\right),\label{kap1}\\
\kappa_2^\infty &=& \frac{2}{K_2 z_B (K_2)} \left(1-e^{-\frac{K_2z_B(K_2)}{2}}\right)e^{-\int_{0}^\infty W_1(z)dz}\,,\nonumber\\ &\simeq & \frac{2}{K_2 z_B (K_2)} \left(1-e^{-\frac{K_2 z_B(K_2)}{2}}\right) e^{-3\pi K_1/8 }\,,\label{kap2}
\eea
where 
\bea
z_{B}(K_i)=2+4K_i^{0.13} e^{-\frac{2.5}{K_i}}\,~{\rm and}~K_i=\frac{(m_D^\dagger m_D)_{ii}}{m^*M_i}
\eea
with $m^*\simeq 10^{-3}$ being the equilibrium neutrino mass.
The frozen out asymmetry  $N_{B-L}^{Df}=\sum_{i}^2\varepsilon_{i}\kappa_{i}^\infty$ then survives down to the low energy with the potential to explain the observed $\eta_B$. The flavoured CP asymmetry parameter is given by\cite{pilaf}
\bea
\varepsilon_{i\alpha}
&=&-\frac{1}{4\pi v^2 h_{ii}}\sum_{j\ne i}\left[ {\rm Im}\{h_{ij}
({m_D^\dagger})_{i\alpha} (m_D)_{\alpha j}\} g({x_{ij})}
+\frac{(1-x_{ij})
{\rm Im}\{{h}_{ji}({m_D^\dagger})_{i\alpha} (m_D)_{\alpha j}\}}
{(1-x_{ij})^2+{{h}_{jj}^2}{(16 \pi^2 v^4)}^{-1}}\right],\label{ncp}
\eea 
where $h_{ij}=(m_D^\dagger m_D)_{ij}$, $x_{ij}=M_j^2/M_i^2$ and $g(x_{ij})$ is given by
\bea
g(x_{ij})=\left[\sqrt{x_{ij}}[1-(1+x_{ij})~{\rm ln}\left(\frac{1+x_{ij}}{x_{ij}}\right)]+\frac{\sqrt{x_{ij}}(1-x_{ij})}
{(1-x_{ij})^2+{{h}_{jj}^2}{(16 \pi^2 v^4)}^{-1}}\right]. \label{gxij}
\eea
Since $h_{ij}$ is a hermitian matrix, when summed over $\alpha$, the second term in Eq.\ref{ncp} vanishes. Using the orthogonal parametrisation for $m_D$ given in Eq.\ref{orth}, the total CP asymmetry parameter (which is relevant in one flavour approximation) can be written as 
\bea
\varepsilon_i &= &-\frac{1}{4\pi v^2}\sum_{\alpha}\frac{ {\rm Im} [M_j\sum_{k k^\prime}\sqrt{m_k m_{k^\prime}}m_k \Omega^*_{ki}\Omega^*_{k^\prime i} U^\dagger_{k^\prime \alpha}U_{\alpha k}]g(x_{ij})}{\sum_{k^{\prime \prime}}m_{k^{\prime \prime}}|\Omega_{k^{\prime\prime} i}|^2} \nonumber\\
&=& -\frac{1}{4\pi v^2}\frac{ M_jg(x_{ij})\sum_{k}m_k^2{\rm Im} [ \Omega^*_{ki} \Omega^*_{ki}  ]}{\sum_{k^{\prime \prime}}m_{k^{\prime \prime}}|\Omega_{k^{\prime\prime} i}|^2},\label{cporth} 
\eea
where $i,j (i\neq j)=1,2$.  In the $N_3$ decoupling limit,  the orthogonal matrices  for NO ($m_1=0$) and IO ($m_3=0$) are given by
\bea
\Omega^{\rm NO}=\begin{pmatrix}
0&0&1\\ \cos\theta &\sin\theta &0\\-\sin\theta & \cos\theta & 0
\end{pmatrix},\hspace{1cm} \Omega^{\rm IO}=\begin{pmatrix}
 \cos\theta &\sin\theta &0\\-\sin\theta & \cos\theta & 0\\0&0&1
\end{pmatrix}, \label{2orth}
\eea
where $\theta=x+iy$ is a complex angle with $x$ and $y$ being real parameters. Using Eq.\ref{2orth} and maximising Eq.\ref{cporth} with respect to $x$ we get\footnote{We choose $x=3\pi/4$ just for a demonstration purpose. However, the overall conclusion drawn is true for all values of $x$.}
\bea
\varepsilon_1^{\rm NO}=-\frac{M_2 g(x_{12})}{4\pi v^2} (m_3-m_2) \tanh 2y,\label{en1} \\
\varepsilon_2^{\rm NO}=\frac{M_1 g(x_{21})}{4\pi v^2} (m_3-m_2) \tanh 2y 
\eea
and
\bea
\varepsilon_1^{\rm IO}=-\frac{M_2 g(x_{12})}{4\pi v^2} (m_2-m_1) \tanh 2y, \\
\varepsilon_2^{\rm IO}=\frac{M_1 g(x_{21})}{4\pi v^2} (m_2-m_1) \tanh 2y.\label{ei2}
\eea
Before an explicit evaluation of the  CP asymmetry parameters, we would like to emphasise on the following: The quantity 
\bea
\gamma_i=\sum_j |\Omega^2_{ij}|\geq 1\label{boost}
\eea
accounts for the fractional contribution of the heavy $M_j$ states to a particular light neutrino $m_i$, and thus it can be treated as a measure of fine-tuning in the seesaw formula\cite{haar}. Since $\Omega$ belongs to $SO(3,\mathbb{C})$, it is isomorphic to the Lorentz group and can be factorized as 
\bea
\Omega = \Omega^{\rm rotation}\Omega^{\rm Boost}.
\eea
Using Eq.\ref{orth} and Eq.\ref{states} one can derive a transformation relation between the states produced by the RH neutrinos ($\ket{\ell_j}$) and the light neutrinos states ($\ket{\tilde{\ell}_i}$) as
\bea
\ket{\ell_j} =B_{ji}\ket{\tilde{\ell}_i},
\eea
where the bridging matrix $B_{ij}$, first introduced in Ref.\cite{haar} relates the heavy and the light states in general with a non-orthonormal transformation  and is related to the orthogonal matrix as 
\bea
B_{ji}=\frac{\sqrt{m_i}\Omega_{ji}}{\sqrt{m_k|\Omega_{kj}|^2}}.
\eea
\begin{figure}
\includegraphics[scale=.38]{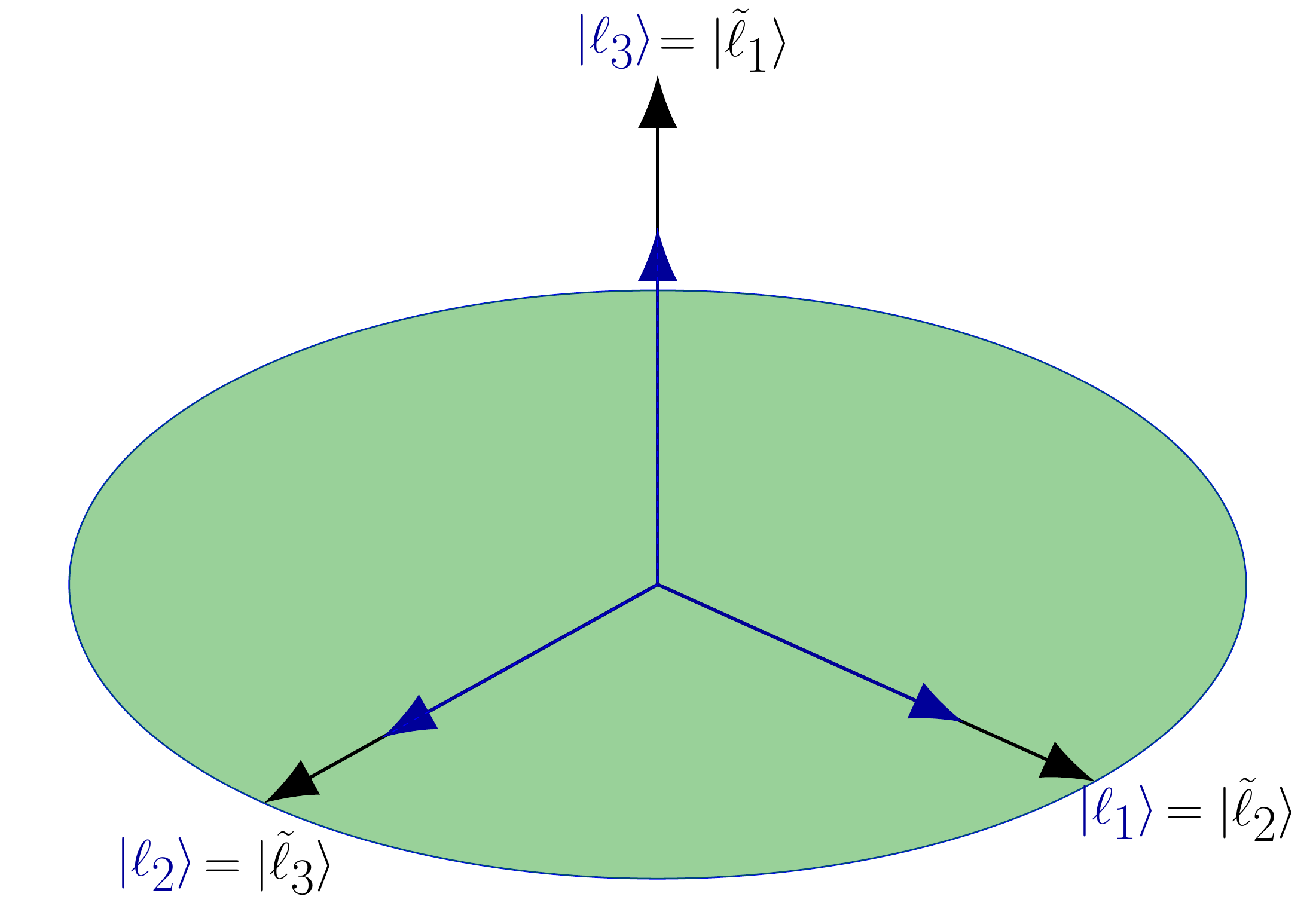}
\caption{Pictorial representation of seesaw models with no fine-tuning, i.e., the states produced by the heavy neutrinos coincide with the orthonormal basis of the light neutrino states.}\label{bases}
\end{figure}
For a choice of the orthogonal matrix $\Omega\equiv P ~(\rm permutation~ matrix)$ which does not correspond to any fine-tuning (a particular heavy neutrino contributes to a particular light neutrino\cite{form,manimala}), e.g.,
\bea
\Omega^{\rm NO}=\begin{pmatrix}
0&0&1\\ 1 &0 &0\\0 & 1 & 0
\end{pmatrix},
\eea
the heavy and the light states coincide as shown in  Fig.\ref{bases}. However for a general  orthogonal matrix (cf.  Eq.\ref{2orth}) which can be factorised as
\bea
\Omega^{\rm NO}=\begin{pmatrix}
0&0&1\\ \cos x &\sin x &0\\-\sin x & \cos x & 0
\end{pmatrix} \begin{pmatrix}
 \cosh y &i \sinh y &0\\-i \sinh y & \cosh y & 0\\0&0&1
\end{pmatrix},
\eea
the orthonormality in the heavy states does not hold unless  $x,y=0$. Due to the presence of the boost matrix, the heavy states are in general strongly non-orthonormal. Using Eq.\ref{boost} the fine-tuning (boost) parameters can be  calculated as 
\bea
\gamma_2=\gamma_3\equiv\gamma=\cosh 2y.
\eea

Thus any non-zero value of $y$ will correspond to a certain level of fine-tuning in the seesaw formula. In fact, for one flavour leptogenesis at least this the case to obtain non-zero CP asymmetry (cf. Eq.\ref{en1}).  Coming back to the discussion of the CP asymmetry parameters, the function $g(x_{12})\propto M_1/M_2$ and $M_1g(x_{21})\propto M_1^2/M_2$. Now e.g., in Eq.\ref{en1}, taking $m_3-m_2\sim 0.1 $ eV, $\tanh 2y\sim 1$ and $\kappa_1^\infty\sim 10^{-2}$ it is evident that one needs $M_1\sim  10^{9}$ GeV to generate $N_{B-L}\sim 10^{-8}$. The contribution to the asymmetry from $N_2$ is negligible since  the CP asymmetry parameter is suppressed by a factor $M_1/M_2$ and the efficiency factor $\kappa_2^\infty$ gets an exponential suppression by $N_1$-washout (cf. Eq.\ref{kap2}). Thus a two RH seesaw model with $M_1<10^9$ GeV and $M_2\gg M_1$ does not lead to successful leptogenesis from RH neutrino decays. We now try to understand  whether the asymmetry  $N_{B-L}^{G0}$ which is generated gravitationally leads to successful leptogenesis. Using the orthogonal parametrisation of $m_D$ in Eq.\ref{orth}, for $M_2\gg M_1$ the asymmetry $N_{B-L}^{G0}$ in Eq.\ref{gl1} can be written as
\bea
N_{B-L}^{eq}=\frac{\pi^2 \dot{R}}{36 (4\pi v)^4}\frac{\sum_k m_k^2 {\rm Im\left[ \Omega_{k1}^*\Omega_{k1}^*\right]}}{\xi(3) T} \frac{M_2^2}{M_1^2}{\rm ln \left(\frac{M_2^2}{M_1^2}\right)}\label{nblg0}
\eea
with $\dot{R}$ as 
\bea
\dot{R}=\sqrt{3}\sigma^{3/2}(1-3 \omega)(1+ \omega) \frac{T^6}{M_{Pl}^3},
\eea
where $\sigma=\pi^2 g^*/30 $, $M_{Pl}\sim 2.4\times 10^{18}$  GeV and we opted for the $p=1$ solution for which the asymmetry gets enhanced  hierarchically\cite{rg9}. A non-zero value of $\dot{R}$ in radiation domination can be obtained by considering so called trace-anomaly in the gauge sector allowing $1-3\omega\simeq 0.1$\cite{tra}. In a seesaw model, the main obstacle to any pre-existing asymmetry to survive down to the Electroweak scale is the washout processes involving lepton number violating $N_i$-interactions. First we consider $M_2 \gg T_0\gg M_1$. Therefore, $N_{B-L}^{G0}$ will face a washout at $T\sim M_1$ (by a factor $e^{-\frac{3\pi}{8}K_1}$)\cite{lu1,lu2}. The final  asymmetry\footnote{This expression is quite robust and perfectly reproduces numerical results as shown by the black dashed line in Fig.\ref{fig2},  that matches the final value of the red solid line.}  is then given by (cf. Eq.\ref{di})
\bea
N_{B-L}^{Gf}=\mathcal{E}_{\Delta L=2}N_{B-L}^{eq}\mathcal{D}(f_{i})=N_{B-L}^{G0}\mathcal{D}(f_{i})=N_{B-L}^{G0}e^{-\frac{3\pi}{8}K_1},\label{unfa}
\eea 
\begin{figure}
\includegraphics[scale=.4]{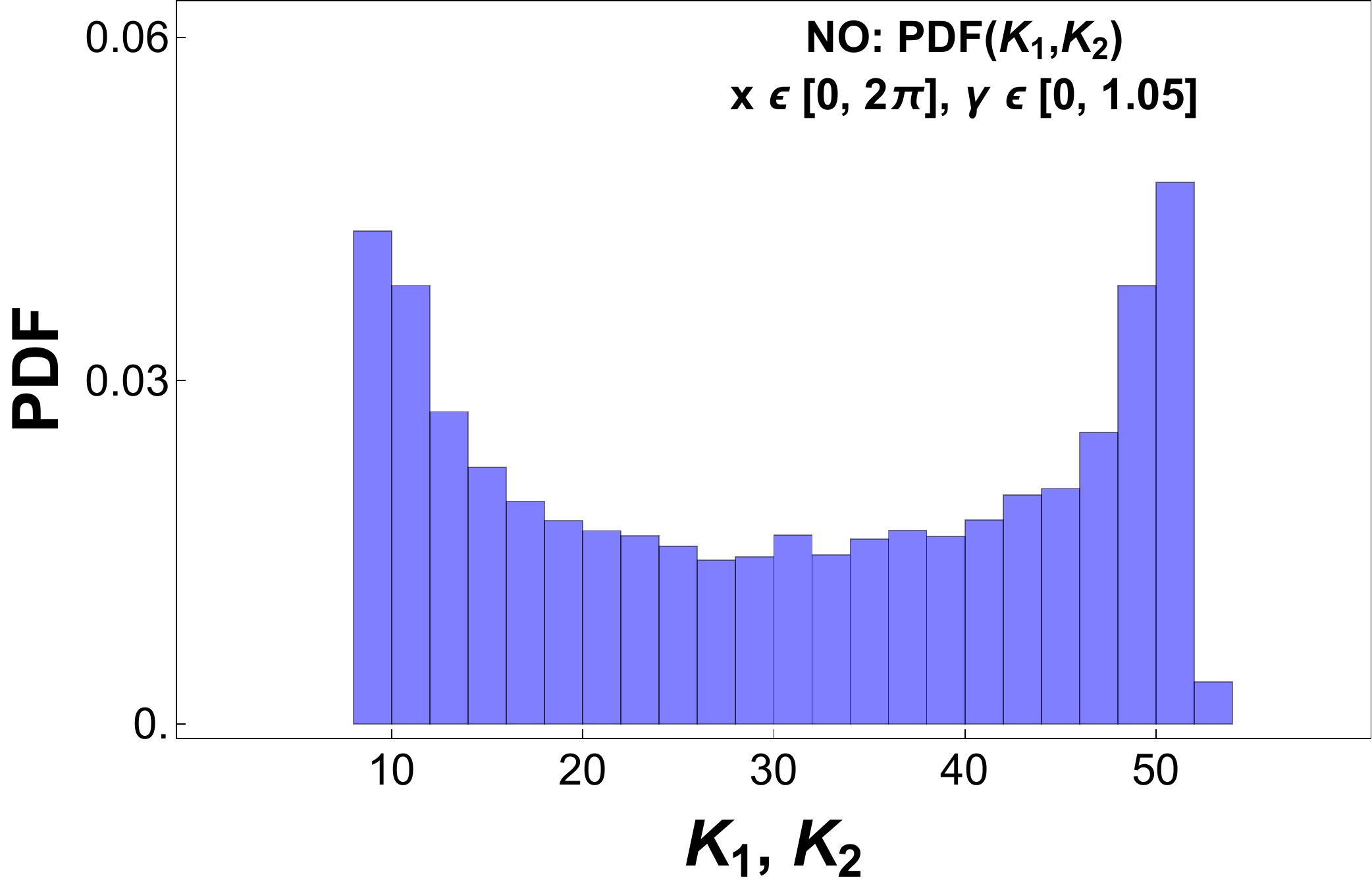}\includegraphics[scale=.5]{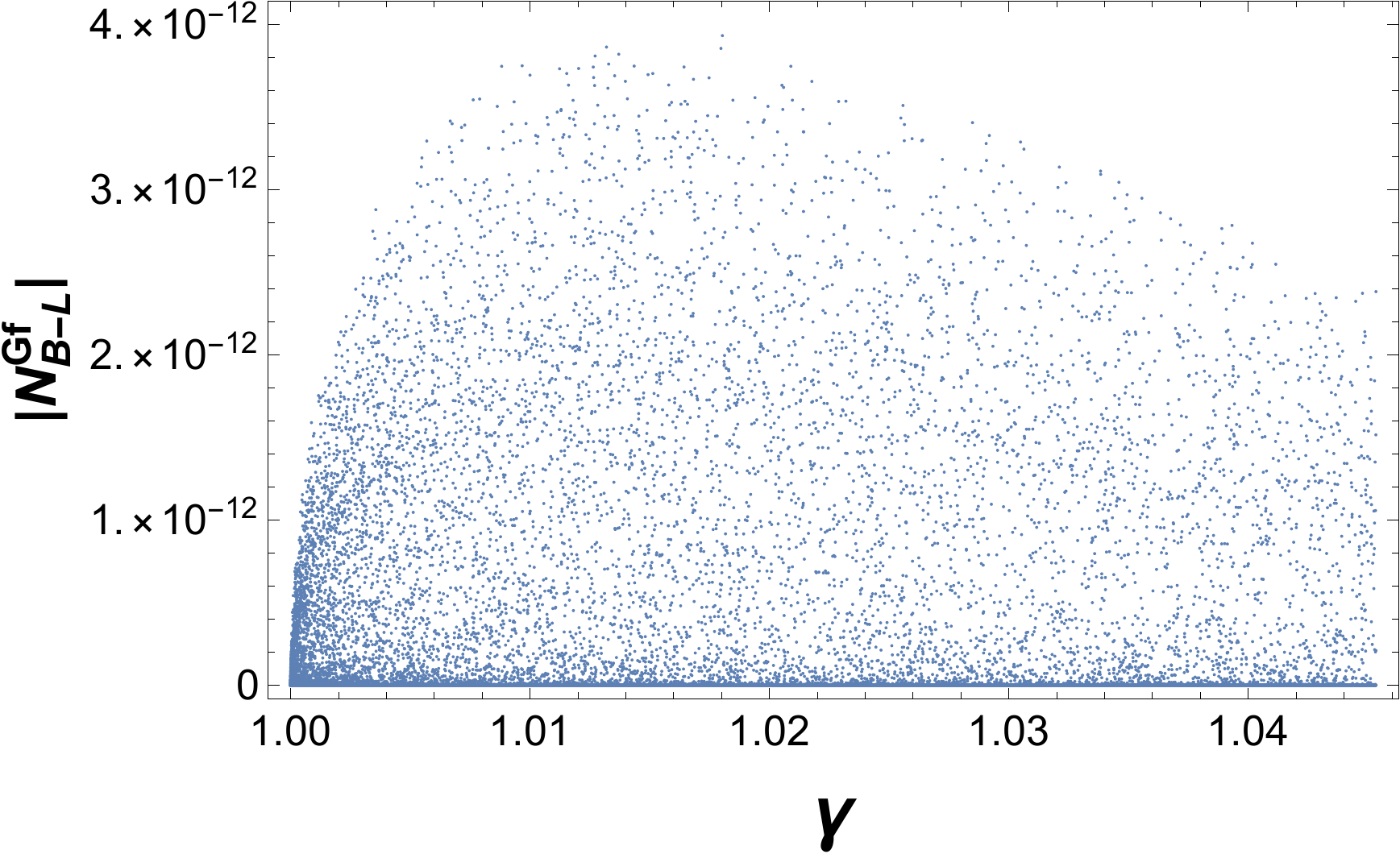}
\caption{Left: Distribution of the total decay parameters ($K_{1,2}$). Right: Magnitude of the gravitational asymmetry after $N_1$-washout.}\label{k1pk2}
\end{figure}
where the decay parameter $K_i$ can be expressed in terms of the orthogonal matrix as
\bea
K_i=\frac{1}{m^*}\sum_k m_k |\Omega_{ki}|^2\label{totdecay}
\eea 
and $\mathcal{E}_{\Delta L=2}$ is the overall dilution factor due to the $\Delta L=2$ processes.  In the left panel of Fig.\ref{k1pk2}, we show a distribution of the decay parameters for minimally fine-tuned ($\simeq 5\%$) seesaw models by considering $x~\epsilon~[0,2\pi]$ and $\gamma ~\epsilon ~[0,1.05]$.  It is evident that due to the large values of the decay parameters the asymmetry $N_{B-L}^{G0}$ gets  washed out strongly and at $z_{B1}$ one finds negligible value of $N_{B-L}^{Gf}$ as shown in the right panel of Fig.\ref{k1pk2}. The plot has been generated for a normal light neutrino mass ordering with a benchmark value of $M_1=10^7$ GeV, $M_2=10^{16}$  GeV, $\mathcal{E}_{\Delta L=2}\sim \mathcal{O}(1)$ (for a realistic flavour case we shall properly deal with $\mathcal{E}_{\Delta L=2}$) and considering the weak gravity condition $T\leq \sqrt{M_1 M_{\rm Pl}}$\cite{rg10}\footnote{In principle, there could be  another condition, namely the low energy condition \cite{rg10}: $z(=M_1/T)\geq (10^{-2} M_1/M_{Pl})^{1/3}$ which will increase our chosen initial value of $z$ by an order of magnitude. Therefore, though we do sacrifice the strict validity of the effective Lagrangian, the point at which it becomes untrustworthy requires a more precise method of calculation. In any case, the gravitational effects go to zero in a dynamically controlled way for temperature well in excess
of the validity of effective Lagrangian\cite{leff1,leff2,leff3}. A more detail discussion in this issue can be found in \cite{rg10}. }. This  conclusion is true also for the strongly boosted seesaw systems ($\gamma\gg 1$) as well as inverted mass ordering.  The case of $T_0\gg M_2\gg M_1$ is more severe. In that case, due to a cascade washout ($\mathcal{D}\sim e^{-(K_1+K_2)}$) at $T_i\sim M_i$, one obtains final asymmetry $\mathcal{O}(10^{-30})$. Thus in the unflavoured regime, with the mass spectrum $M_2\gg M_1$, gravitational leptogenesis fails to explain the observed $\eta_B$. Whilst it is well known that for the discussed spectrum of masses, successful leptogenesis from decays is not possible even if one considers flavour effects (though the washouts get reduced, one simply does not have sufficient CP asymmetry from both the RH neutrinos)\cite{2rh1,2rh2}, consideration of flavour effects in gravitational leptogenesis in this context requires investigation. This is the main objective of this work. The basic idea is that when one considers flavour effects, instead of total decay parameters $K_i$, flavoured decay parameters $K_{i\alpha}$ appear in the exponential washout (cf. Eq.\ref{unfa}). We will see in the next section that $K_{i\alpha}=P_{i\alpha}K_i$ with $P_{i\alpha}<1$ is the probability of a flavour state $\ket{\ell_\alpha}$ being in the state $\ket{\ell_i}$ associated to the heavy neutrinos. Thus $K_{i\alpha}$ is always weaker than $K_i$ and the washout effects get reduced which in turn enhance the probability for $N_{B-L}^{Gf}$ being sizeable enough to be consistent with $\eta_{CMB}$.

\section{Flavour effect and successful gravitational leptogenesis}\label{s3}
Depending on the mass of the RH neutrinos, flavour effects play a crucial role in the computation related to leptogenesis. The one flavour regime (1FR) is typically characterised by $M_i >10^{12}$ GeV where all the charged lepton flavours are out of equilibrium, and thus the lepton doublet $\ket{\ell_i}$ produced by the decay of the RH neutrinos or other external sources can be written as a coherent superposition of the corresponding flavour states $\ket{\ell_{\alpha}}$ as given in Eq.\ref{coh1} and Eq.\ref{coh2}.  Since there is hardly any interaction to break the coherence of the quantum states  e.g.,  before it inversely decays to $N_i$ ($N_i$-washout), the asymmetry is produced along the direction of $\ket{\ell_i}$(or $\ket{\bar{\ell}_i}$) in the flavour space. However, this is not the case when $M_i<10^{12}$ GeV, since below this scale, the $\tau$ charged lepton flavour comes into equilibrium and breaks the coherence of $\ket{\ell_i}$ states ($\tau$ component gets measured\cite{fl1}). Thus the relevant flavours that take part in the washout processes are the flavour $\tau$ and the coherent superposition of the flavours $e$ and $\mu$ - this is so called the two flavour regime (2FR)\cite{fl1,fl5}. Similarly when $M_i<10^{9}$ GeV, fast $\mu$ flavour interactions break the coherence of $e$ and $\mu$, therefore, one resolves all the three  flavours (3FR).  The flavour effects are taken into account by defining the branching ratios into individual flavours as $P_{i\alpha}=|\mathcal{A}_{i\alpha}|^2$ and $\bar{P}_{i\alpha}=|\bar{\mathcal{A}}_{i\alpha}|^2$. As a result, the decays (and hence the inverse decays, $\Gamma_i^{\rm ID}=\Gamma_i\frac{N_i^{\rm eq}(z)}{N_\ell^{\rm eq}}$) into individual flavours could be written as $\Gamma_{i\alpha}\equiv P_{i\alpha}$ $\Gamma_i$ and $\bar{\Gamma}_{i\alpha}\equiv\bar{P}_{i\alpha}\bar{\Gamma}_i$ with $\sum_{\alpha}(P_{i\alpha},\bar{P}_{i\alpha})=1$. It is  convenient also to introduce the flavoured decay parameter $K_{i\alpha}$ given by 
 \bea
 K_{i\alpha}=\frac{\Gamma_{i\alpha}+\bar{\Gamma}_{i\alpha}}{H(T=M_i)}\simeq \frac{ P_{i\alpha}^0(\Gamma_i+\bar{\Gamma}_i)}{H(T=M_i)}\equiv P_{i\alpha}^0 K_i\equiv\frac{|m_{D_{i\alpha}}|^2}{M_i m^*}\label{fldecay}
 \eea
which in terms of the orthogonal matrix can be re-expressed as
\bea
K_{i\alpha}=\frac{1}{m^*}\left | \sum_k U_{\alpha k}\sqrt{m_k}\Omega_{ki}\right |^2.\label{fdp}
\eea
In the washout processes what matters is thus the flavoured decay parameters, e.g, the efficiency factor in Eq.\ref{effi} can now be generalised to
\bea
\kappa_{i\alpha} (z)=-\frac{4}{3}\int_{z_{\rm in}}^z \frac{dN_{N_i}}{dz^\prime}e^{-\sum_{j}\int_{z^\prime}^z P^0_{j\alpha}W_{j}^{\rm ID}(z^{\prime\prime})dz^{\prime\prime}}dz^\prime\,\label{effifl}
\eea
and the final value of any pre-existing asymmetry, e.g.,  $N_{B-L}^{Gf}$ can be written\footnote{We do not consider flavour effects on non-resonant $\Delta L=2$ processes since as we will see, bulk of the allowed solutions correspond to  $\Delta L=2$ processes which are weaker and can be achieved in one flavour approximation. However in the case of precision calculation,  it should be taken into account. } as 
\bea
N_{B-L}^{Gf}=\sum_\alpha p_{\alpha G} N_{B-L}^{G0} e^{-\sum_{j}\int_{0}^\infty P^0_{j\alpha}W_{j}^{\rm ID}(z^{\prime\prime})dz^{\prime\prime}}dz^\prime\,,\label{grava}
\eea
where $p_{\alpha G}$ is the probability for the $\ket{\ell_G}$ states being in the flavour $\alpha$. We assume production of  gravitational asymmetry happens in the unflavoured regime (flavour blind) and conservatively consider $p_{\alpha G}=1/3$ through out\cite{lu1}. As mentioned earlier, we intend to discuss the mass spectrum a)  $10^9 {\rm GeV}\ll M_2\ll 10^{12} {\rm GeV}\lesssim T_0$, and b) $ 10^{12} {\rm GeV}\lesssim T_0\ll M_2$, $M_1\ll 10^9$  GeV, therefore, Eq.\ref{effifl} is irrelevant in our discussion since this is the flavoured efficiency factors for lepton asymmetry produced by RH neutrino decays which for the chosen spectrum of RH masses, does not contribute significantly to the final $\eta_B$. Thus Eq.\ref{grava} is the key equation for the entire analysis. As an aside, let's have a technical remark regarding the computation. We are considering strong flavour effect so that throughout the washout phases, charged lepton interactions dominate over the washout processes and the lepton system is completely incoherent in relevant flavour. Thus the formulae we use are the solutions of flavour diagonal Boltzmann Equations\cite{lu2,fl5}. Otherwise, one has to solve exact density matrix equations which also take care of coherence among the flavour states. Strong flavour effect can be implemented by the condition
\bea
W_i^{\rm max}(z_i\approx 1)<F_\alpha
\eea
with $W_i$ is the washout parameter and $F_\alpha=\frac{\Gamma_\alpha}{H z_i}$ is the rate of charged lepton interaction. Taking $W_i\equiv W_i^{\rm ID}$ as\cite{bari}
\bea
W_i^{\rm ID}=\frac{1}{4}K_iz_i^2\sqrt{1+\frac{\pi}{2}z_i}e^{-z_i}
\eea
at $z_i\approx 1$ one arrives 
\bea
\frac{3K_i}{20}<F_\alpha \label{strong}
\eea
where $F_\tau=10^{12}/M_i$ and $F_\mu=10^{9}/M_i$.  Eq.\ref{strong} when combined with Eq.\ref{totdecay}, translates into a condition on the boost parameter as
\bea
\gamma\equiv \cosh 2y \lesssim \frac{F_\alpha}{75}\left(\sum_i m_i/{\rm eV}\right)^{-1}. \label{stfl}
\eea
 This is the restriction one has to impose on the parameter space for the strong flavour effect to be strictly valid. Though in the end, we will see that Eq.\ref{stfl} has only a very mild effect on the parameter space pertaining to successful leptogenesis.
 
Having set up the basic formalism and technicalities of flavour effects in leptogenesis we now turn to the detailed analysis of the flavour effect in gravitational leptogenesis with the concerned spectrum of RH masses.\\

\begin{figure}
\includegraphics[scale=.5]{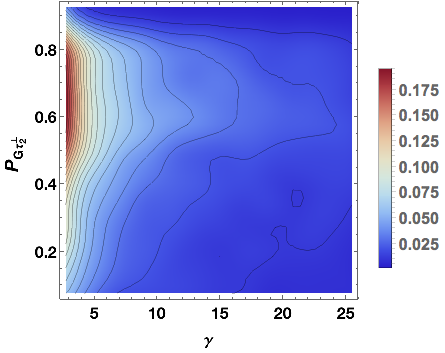}\includegraphics[scale=.4]{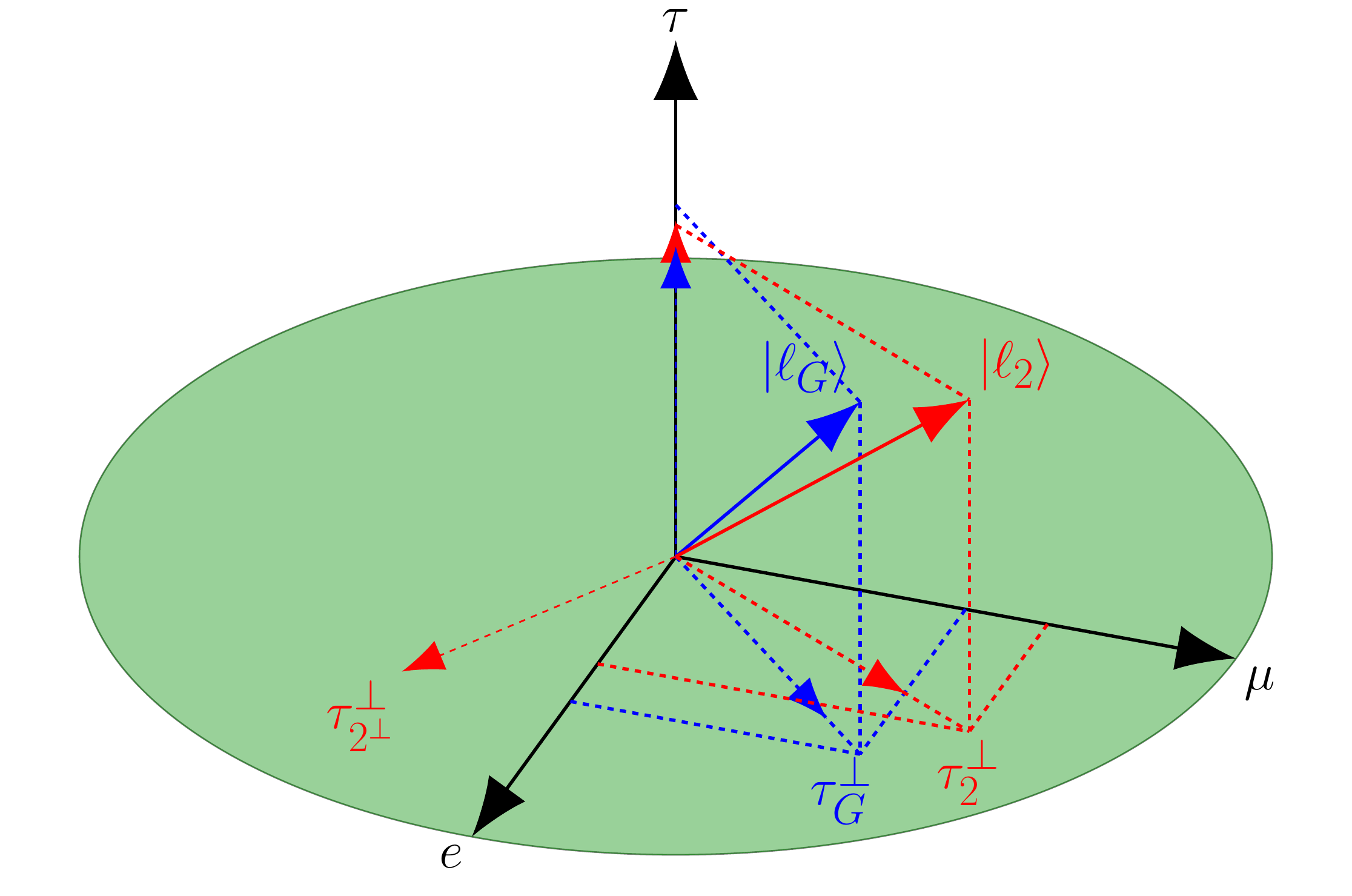}
\caption{Left: Distribution of the overlap probability $p_{G\tau_2^\perp}$. Right: A visual representation of relevant flavour directions pertinent to  leptogenesis for the mass spectrum $10^9 {\rm GeV}\ll M_2\ll 10^{12}$ \rm GeV, $M_1\ll 10^9$ \rm GeV.  }\label{2flp}
\end{figure}

 \paragraph{$10^9 {\rm GeV}\ll M_2\ll 10^{12}$ \rm GeV $\lesssim T_0$, $M_1\ll 10^9$ \rm GeV :} In this case, $N_2$-washout happens in the two flavour regime and washout by $N_1$ acts in the three flavour regime. We can write  $N_{B-L}^{G0} (T \gg M_2)$  as a sum of two components:
\bea
N_{B-L}^{G0}\equiv N_{B-L}^{G0\tau}+ N_{B-L}^{G0\tau^\perp} = p^0_{\tau G}N_{B-L}^{G0}+(1-p^0_{\tau G})N_{B-L}^{G0},
\eea
where $ p^0_{\tau G}$ is the probability of $N_{B-L}^{G0}$ being in the direction of $\tau$. Consequently, $ 1-p^0_{\tau G}$ is the probability in the $\tau^\perp$ direction.
\subparagraph{Flavour coincidence:} In this subsection, we present the basic idea of the flavour effects assuming $\tau$ and $\tau^\perp$ directions of $\ket{\ell_{G}}$ coincide with that of $\ket{\ell_2}$ (which is in general not true, see Fig.\ref{2flp}). There are now two stages of washout. The first one is at $T\sim M_2$ and the final one is at $T\sim M_1$. After the end of the first phase of washout, i.e., after  $N_2$-interactions go out of equilibrium at $(T\sim z_{B2})$, the combined contribution of $\tau$ and $\tau^\perp_G$ component to the final asymmetry can be written as
\bea
N_{B-L}^{G1}&=&N_{B-L}^{G1\tau}+N_{B-L}^{G1\tau^\perp}=p^0_{\tau G}e^{-3\pi(K_{2\tau})/8}N_{B-L}^{G0}+(1-p^0_{\tau G})e^{-3\pi(K_{2e}+K_{2\mu})/8}N_{B-L}^{G0},
\eea 
where $N_{B-L}^{Gi}$ is the frozen out asymmetry after $i$th stage of washout. The frozen out asymmetries in each of the components will then face  $N_1$-washout  and at $z_{B1}$ the components of final unwashed asymmetry can be written as
\bea
N_{B-L}^{G2\tau}=N_{B-L}^{G1\tau}e^{-3\pi(K_{1\tau})/8} =p^0_{\tau G}e^{-3\pi(K_{1\tau}+K_{2\tau})/8}N_{B-L}^{G0},
\eea
\bea
N_{B-L}^{G2\mu}=p_{\mu \tau_2^\perp}^0N_{B-L}^{G1\tau^\perp}e^{-3\pi(K_{1\mu})/8} =p_{\mu\tau_2^\perp}^0(1-p^0_{\tau G})e^{-3\pi(K_{2e}+K_{2\mu}+K_{1\mu})/8}N_{B-L}^{G0},
\eea
\bea
N_{B-L}^{G2e}=p_{e\tau_2^\perp}^0N_{B-L}^{G1\tau^\perp}e^{-3\pi(K_{1e})/8} =p_{e\tau_2^\perp}^0(1-p^0_{\tau G})e^{-3\pi(K_{2e}+K_{2\mu}+K_{1e})/8}N_{B-L}^{G0}
\eea
where the probabilities $p_{\alpha\tau_2^\perp}^0$ are given by
\bea
p_{\alpha\tau_2^\perp}^0=\frac{p^0_{2\alpha}}{\sum_\alpha p^0_{2\alpha}}=\frac{K^0_{2\alpha}}{\sum_\alpha K^0_{2\alpha}} ~~{\rm with }~~ \alpha=e,\mu.
\eea
The final asymmetry is then given by
\bea
N_{B-L}^{Gf}=\sum_\alpha N_{B-L}^{G2\alpha}~{\rm with}~\alpha=e,\mu,\tau.
\eea
\subparagraph{Projection dominance:} In the previous discussion  to have an overall idea of the washout processes we assume the states $\ket{\ell_{\tau_G^\perp}}$ and $\ket{\ell_{\tau_2^\perp}}$ in the $e-\mu$ plane share a common direction, i.e., $p_{G\tau_2^\perp}\equiv|\braket{\ell_{\tau_G^\perp}|\ell_{\tau_2^\perp}}|^2=1$. However, this is in general not true. Assuming flavour blind production of gravitational asymmetry, the probability $p_{G\tau_2^\perp}$ can be calculated as
\bea
p_{G\tau_2^\perp}=\frac{1}{2}\frac{K_2}{K_{2\tau^\perp}}\frac{\left | \sum_k \sqrt{ m_k} U_{e k}^* \Omega_{k 2}^*+\sum_k \sqrt{ m_k} U_{\mu k}^* \Omega_{k 2}^*\right |^2}{\sum_k m_k |\Omega_{k 2}|^2}.
\eea
%
In Fig.\ref{2flp} we plot a distribution of $p_{G\tau_2^\perp}$ with the $\gamma$. It is evident that $p_{G\tau_2^\perp}$ can have any values ranging from $0-1$. In fact, the most probable values are clustered around $p_{G\tau_2^\perp}\simeq 0.7$. Thus component of $N_{B-L}^{G0\tau^\perp}$ which is in the direction of $\ket{\ell_{\tau_2^\perp}}$ will be washed out by $N_2$ interactions but the component which is orthogonal to $\ket{\ell_{\tau_2^\perp}}$ will escape $N_2$-washout. This is so called the projection effect first introduced in Ref.\cite{nardi} and then studied in detail e.g., in Refs.\cite{lu2,fl5}. The final baryon asymmetry at $z_{B2}$ can now be written as
\bea
N_{B-L}^{G1\tau}&=&p^0_{\tau G}e^{-3\pi(K_{2\tau})/8}N_{B-L}^{G0},\\
N_{B-L}^{G1\tau^\perp}&=&N_{B-L}^{G1\tau_2^\perp}+N_{B-L}^{G1\tau_{2^\perp}^\perp},\label{n2w}
\eea
where 
\bea
N_{B-L}^{G1\tau_2^\perp}&=&(1-p^0_{\tau G})p_{G\tau_2^\perp}e^{-3\pi(K_{2e}+K_{2\mu})/8}N_{B-L}^{G0},\\
N_{B-L}^{G1\tau_{2^\perp}^\perp}&=&(1-p^0_{\tau G})(1-p_{G\tau_2^\perp})N_{B-L}^{G0}
\eea
Now proceeding in the same way as the flavour coincidence case, after the $N_1$-washout, the asymmetries in each flavour can be written as
\bea
N_{B-L}^{G2e}&=&(1-p^0_{\tau G})e^{-3\pi(K_{1e})/8}\left[p_{G\tau_2^\perp} p_{e\tau_2^\perp}^0e^{-3\pi(K_{2e}+K_{2\mu})/8}+(1-p_{e\tau_2^\perp}^0)(1-p_{G\tau_2^\perp})\right]N_{B-L}^{G0}, \nonumber \\
N_{B-L}^{G2\mu}&=&(1-p^0_{\tau G})e^{-3\pi(K_{1\mu})/8}\left[p_{G\tau_2^\perp} p_{\mu\tau_2^\perp}^0e^{-3\pi(K_{2e}+K_{2\mu})/8}+(1-p_{\mu\tau_2^\perp}^0)(1-p_{G\tau_2^\perp})\right]N_{B-L}^{G0}, \nonumber \\
N_{B-L}^{G2\tau}&=&p^0_{\tau G}e^{-3\pi(K_{2\tau}+K_{2\tau})/8}N_{B-L}^{G0}.\label{2fl}
\eea
%
\begin{figure}
\includegraphics[scale=.55]{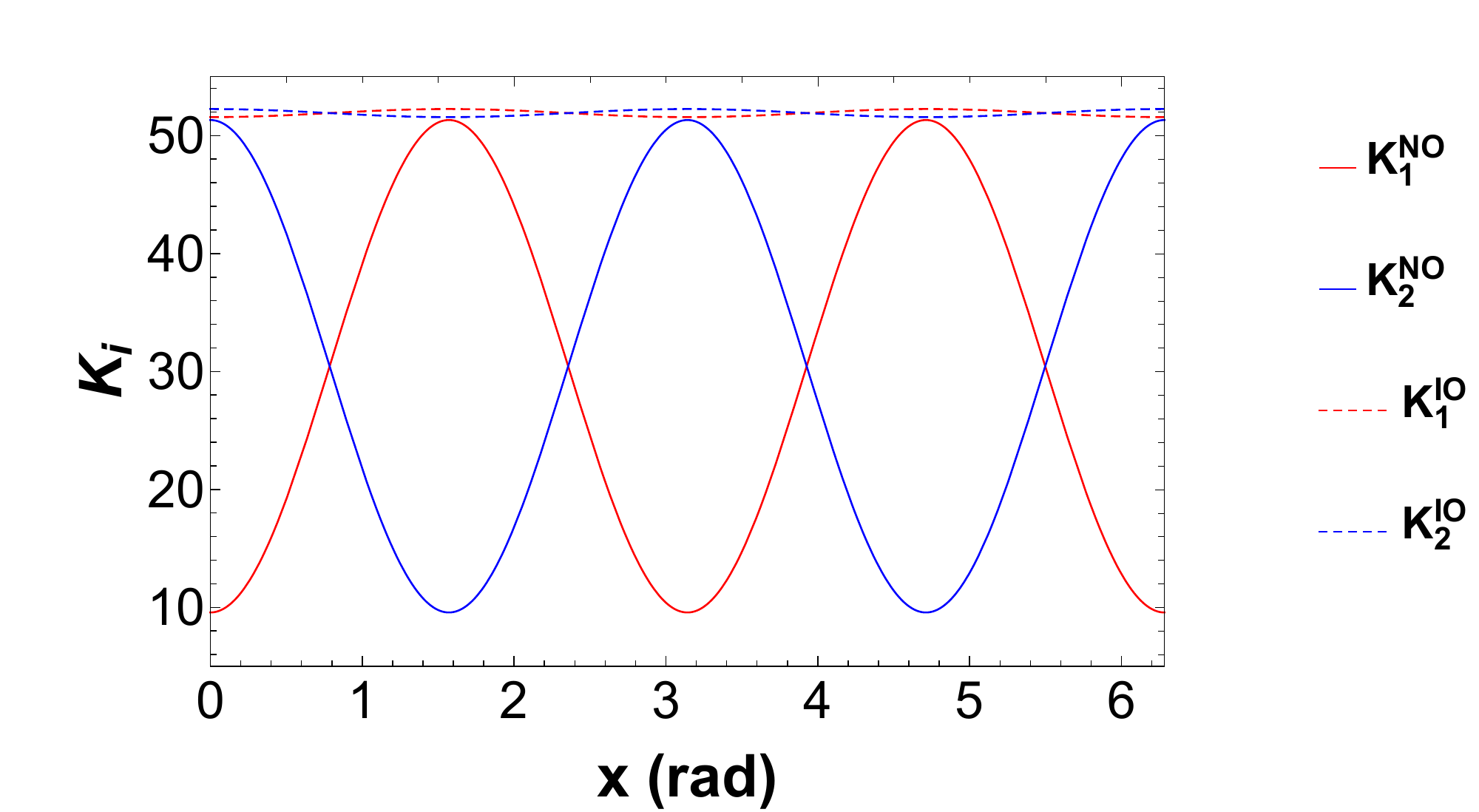}
\caption{ Unflavoured decay parameters in two RH neutrino seesaw models with $\gamma=1.05$.  }\label{k1k2}
\end{figure}

For $p_{G\tau_2^\perp}=1$ we recover the formulae for flavour asymmetries presented in the flavour coincidence case. Note that, the second term (appears due to projection effect) in the first two equations in Eq.\ref{2fl} are dominating since compare to the other terms they escape $N_2$-washout and face washout by $N_1$ only at the second stage. We shall present numerical analysis only for the projection dominance since this is a more complete scenario compared to the flavour coincidence case. \\

 b) $ 10^{12} {\rm GeV}\lesssim T_0\ll M_2$, $M_1\ll 10^9$  GeV:  In this case, $N_2$ does not participate in the washout process whereas  the $N_1$-washout happens in the three flavour regime. The  final asymmetry is then given by relatively simple formula 
 \bea
N_{B-L}^{Gf}=N_{B-L}^{G0}\left(p_{eG}e^{-3\pi(K_{1e})/8}+p_{\mu G}e^{-3\pi(K_{1\mu})/8}+p_{\tau G}e^{-3\pi(K_{1\tau})/8}\right).\label{3fl}
\eea 
\begin{figure}
\includegraphics[scale=.37]{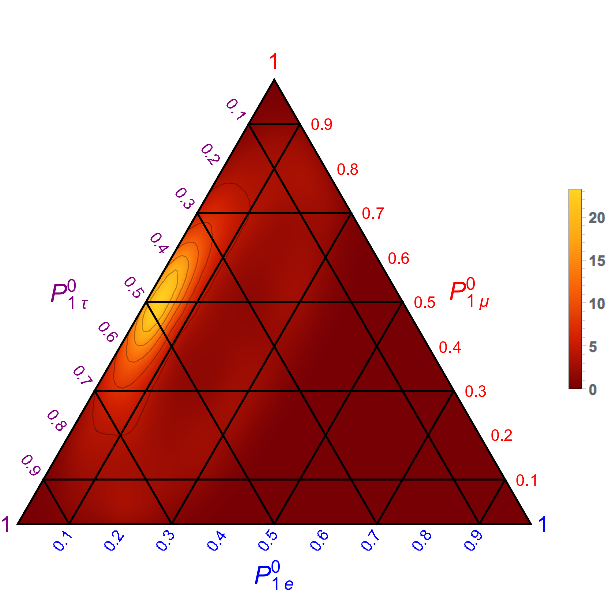}\includegraphics[scale=.37]{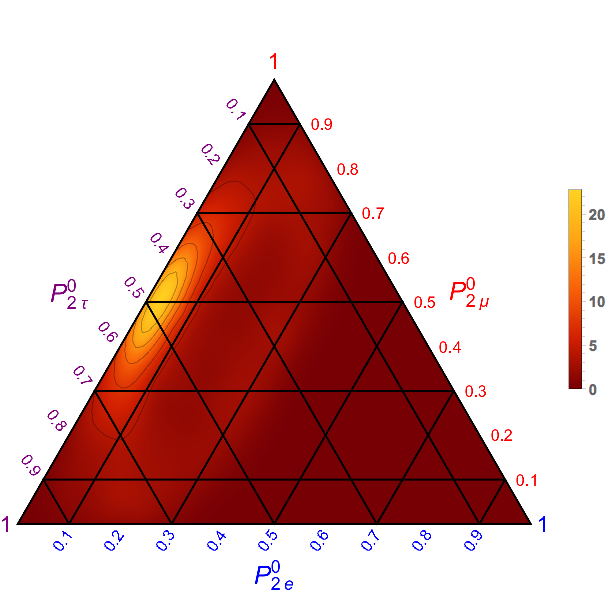}
\includegraphics[scale=.37]{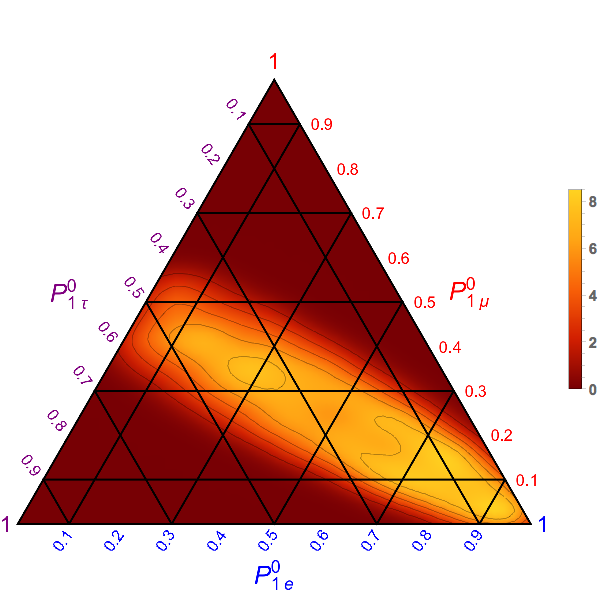}\includegraphics[scale=.37]{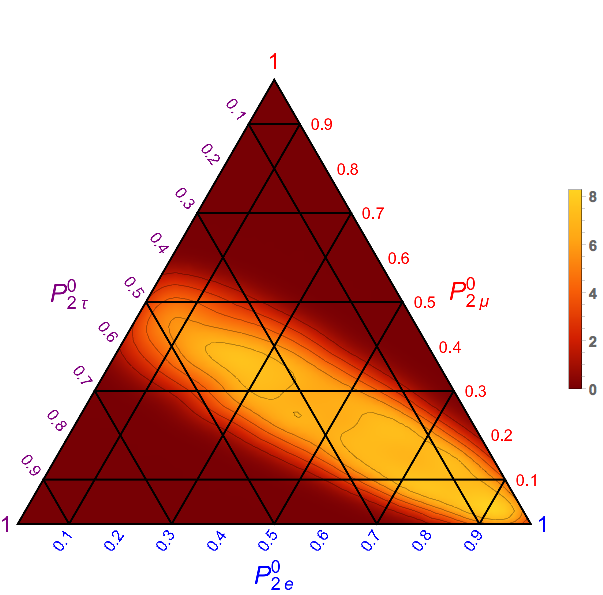}
\caption{Ternary plots of the flavour probabilities. Upper panel: Normal mass ordering. Lower panel: Inverted mass ordering. }\label{tri}
\end{figure}
Given the expression for the final asymmetry in Eq.\ref{2fl} and Eq.\ref{3fl} we  now try to understand quantitatively why in  flavoured case the washout is less. 
First of all, using Eq.\ref{totdecay}, the unflavoured decay parameters can be written as 
\bea
K_1=\frac{1}{2m^*}\left[(m_\alpha-m_\beta) \cos 2x+(m_\alpha+m_\beta)\cosh 2y\right],\\
K_2=\frac{1}{2m^*}\left[-(m_\alpha-m_\beta) \cos 2x+(m_\alpha+m_\beta)\cosh 2y\right],
\eea
where for a normal ordering $\alpha=2,~\beta=3$ and for an inverted ordering  $\alpha=1,~\beta=2$. Smaller values of the unflavoured decay parameters are obtained for $\gamma=\cosh 2y\simeq 1$. For a normal light neutrino mass ordering,  minimum value of $K^{\rm min}_{1,2}=m_2/m^*\equiv \sqrt{\Delta m^2_{12_{\rm min}}}/m^* \simeq 8.4$ is obtained for $x=n \pi$ and $x=(2n+1) \pi/2$ respectively (cf. Fig.\ref{k1k2}). On the other hand for an inverted mass ordering $K^{\rm min}_{1,2}=m_1/m^*\equiv \sqrt{\Delta m^2_{32_{\rm min}}- \Delta m^2_{12_{\rm min}}}/m^* \simeq 49$ is obtained for the same values of $x$. Therefore, in the unflavoured regime, washout ($\sim e^{-49}$) in  inverted mass ordering  is more severe than the washout  ($\sim e^{-8}$) in normal mass ordering. However when flavour effect is included the scenario is quite different. The key physics that is responsible for weakening the strength of $K_{i\alpha}$ is the appearance of light neutrino mixing matrix $U$ in Eq.\ref{fdp}. Using $3\sigma$  neutrino oscillation data, in Fig.\ref{tri} we have shown a model independent triangle quantization of the flavour space. The upper panel corresponds to a normal and the lower panel corresponds to an inverted light neutrino mass ordering. One observes that for normal mass ordering the probability for $K_{ie}=P_{ie}K_i< 1$ is much higher than the other two flavours. This means for the normal light neutrino ordering final asymmetry will be dominated by electron flavour (less washout in the electron flavour). On the other hand for inverted mass ordering the probability of having lower values $K_{i\alpha}$ is quite democratic (though there is a   bias towards $\mu$ and $\tau$ flavours). In any case, the overall information what we obtain from Fig.\ref{tri} is that we can have $K_{i\alpha}\ll 1$ to consider the washout at $T_i\sim M_i$  less significant.   Now the leftover task is to precisely compute $N_{B-L}^{G0}$ considering the effect of $\Delta L=2$ processes which tend to maintain the asymmetry $N_{B-L}^{eq}$ in equilibrium and therefore a dilution of the asymmetry between $z_{in}$ to $z_0$ (where the asymmetry freezes out). 
The frozen out asymmetry $N_{B-L}^{G0}$ can be calculated by solving a simple Boltzmann equation\cite{rg9}\footnote{Recently another term in the B.E has been introduced in Ref.\cite{rg10} which moderates the behaviour of $N_{B-L}$ at ultra-high temperature. However, in this paper we neglect that term for simplicity. } 
\begin{figure}
\includegraphics[scale=.4]{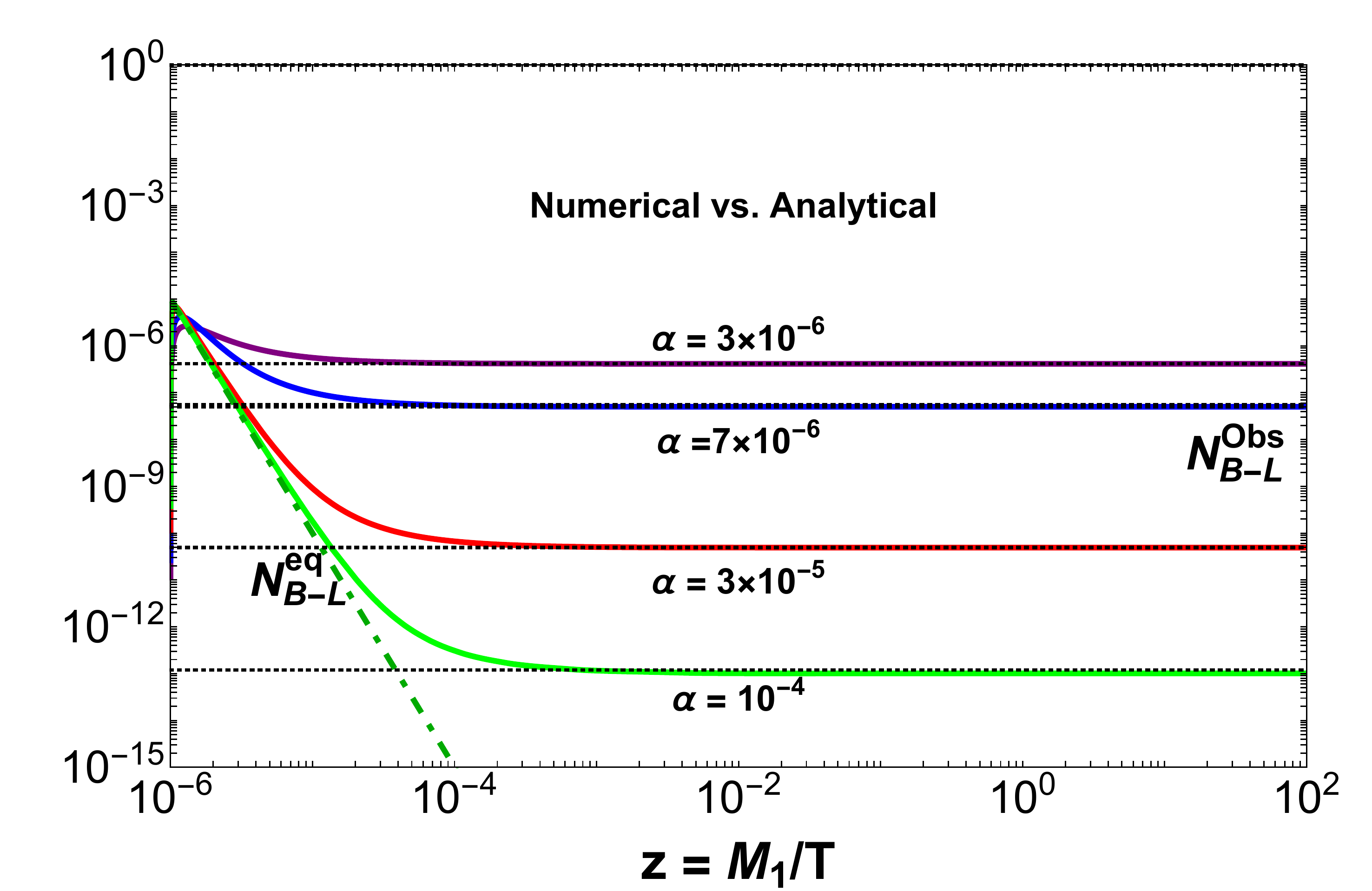}
\caption{A numerical vs. analytical comparison of $N_{B-L}^{G0}$. The coloured lines are numerical solutions and the black dashed lines are analytical yields.}\label{nva}
\end{figure}
\bea
\frac{dN_{B-L}}{dz}=-W_{\Delta L=2}(N_{B-L}-N_{B-L}^{eq}),\label{BE}
\eea
where $W_{\Delta L=2}$ encodes the effect of $\Delta L= 2$ process  involving non-resonant $N_1$-exchange and is given by \cite{pre1,rg10}
\bea
W_{\Delta L=2}(z\ll 1)\simeq \frac{12m^* M_1}{\pi^2 v^2 z^2}\left( \left[\frac{\bar{m}}{m^*}\right]^2+K_1^2\right)~~{\rm with}~~\bar{m}=\sqrt{\sum_i m_i^2}.
\eea
Since we intend to scan the entire parameter space using $3\sigma$ neutrino oscillation data, it is convenient to solve the BE in Eq.\ref{BE} analytically. To this end,
we re-write Eq.\ref{BE} as
\bea
\frac{dN_{B-L}}{dz}=-\frac{\alpha}{z^2}\left(N_{B-L}-\frac{\beta}{z^5}\right),
\eea
where 
\bea
\alpha=\frac{12m^* M_1}{\pi^2 v^2 }\left( \left[\frac{\bar{m}}{m^*}\right]^2+K_1^2\right),~\beta=\frac{\sqrt{3}\sigma^{3/2}M_1^5}{M_{Pl}^3}(1-3\omega)(1+\omega)\mathcal{Y}.
\eea
The parameter  $\mathcal{Y}$ which encodes the CP violation in the theory is given by
\bea
\mathcal{Y}=\frac{\pi^2 }{36 (4\pi v)^4}\frac{\sum_k m_k^2 {\rm Im\left[ \Omega_{k1}^*\Omega_{k1}^*\right]}}{\xi(3) } \frac{M_2^2}{M_1^2}{\rm ln \left(\frac{M_2^2}{M_1^2}\right)}.
\eea 
Starting from a vanishing initial abundance of $N_{B-L}(z)$, for large values of $z$ we find the analytical  solution for $N_{B-L}^{G0}$ as
\bea
N_{B-L}^{G0}=\frac{120\beta}{\alpha^5}\left[1-e^{-\alpha/z_{\rm in}}\right]-\frac{\beta e^{-\alpha/z_{\rm in}}}{\alpha^5}\left[\sum_{n=1}^5\frac{5!}{n!}\left(\frac{\alpha}{z_{\rm in}}\right)^n\right].\label{key}
\eea
Eq.\ref{key} is the key analytical formula  for $N_{B-L}^{G0}$ with a minimum value of $z_{\rm in}^{\rm min}=\sqrt{M_1/M_{Pl}}$ that we use to scan the parameter space.

\section{   parameter space and final results}\label{s4}
\begin{figure}
\includegraphics[scale=.45]{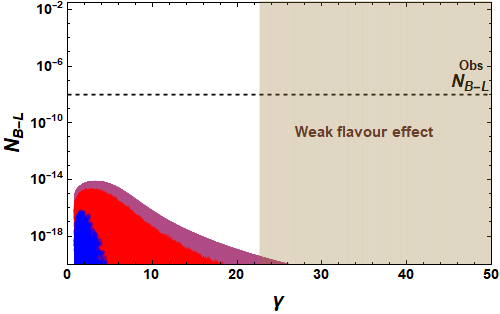}\includegraphics[scale=.45]{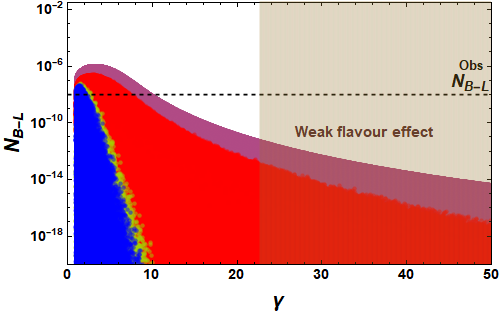}\\
\includegraphics[scale=.45]{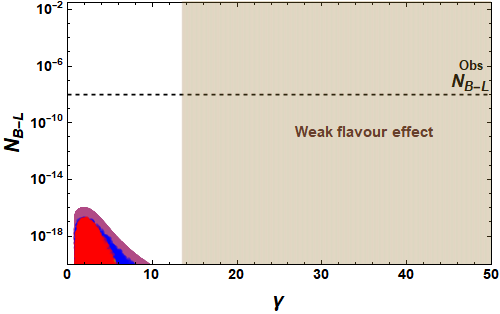}\includegraphics[scale=.45]{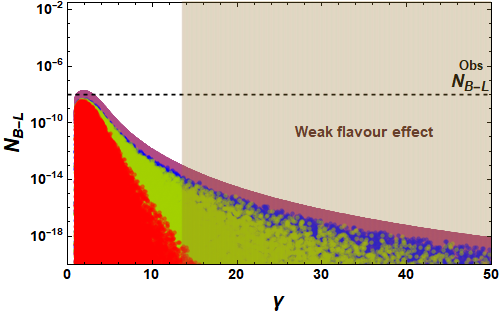}
\caption{Upper panel left: case a) NO: $M_2=10^{12}$ GeV, $M_1=10^7$ GeV, right: case b) NO: $M_2=10^{16}$ GeV, $M_1=10^7$ GeV. Lower panel left: case a) IO: $M_2=10^{12}$ GeV, $M_1=10^7$ GeV, right: case b) IO: $M_2=10^{16}$ GeV, $M_1=10^7$ GeV. All the neutrino oscillation parameters are varied within $3\sigma$ whereas $x$ and $\gamma$ are varied as $x~\varepsilon~[0,2\pi]$ and $\gamma~\varepsilon~[0,50]$.}\label{para}
\end{figure}

First, we compare the final frozen out value of the asymmetry $N_{B-L}^{G0}$ that is obtained from the analytical formula in Eq.\ref{key}  with the numerical solutions of Eq.\ref{BE}. In Fig.\ref{nva} for a benchmark value of $z_{\rm in}=10^{-6}$ and $\beta=10^{-35}$, we show the evolution of the asymmetry (not taking into account the washout at $T_1\sim M_1$) for different values of $\alpha$. The coloured lines are the numerical solutions, and the black dashed lines represent analytically obtained values of $N_{B-L}^{G0}$ which perfectly match the numerical results. It's worth noticing that for large values of $\alpha$, the asymmetry closely tracks its equilibrium value and therefore suffers a late freeze-out which in turn reduces the magnitude of the final asymmetry. We shall see this feature in our final results as well. We can now convincingly use the formula in Eq.\ref{key} along with  Eq.\ref{2fl} and Eq.\ref{3fl} to scan  entire parameter space using $3\sigma$ neutrino oscillation data. In  Fig.\ref{para}, we show our final result. The upper panel contains parameter spaces for normal mass ordering for both the cases (left: case-a, right: case-b) and the lower panel contains the parameter spaces for the inverted mass ordering with the relevant cases ordered in a similar manner as in the upper panel. To generate the figures we fix $M_1=10^7$ GeV for all the cases  and $M_2=10^{16}$ GeV for case-b and $M_2=10^{12} (M_2^{\rm max})$ GeV for case-a (since we are considering $M_2$ to be in the two flavour regime for case-a).  All the other parameters are randomly varied, i.e., the neutrino oscillation parameters are randomly generated following a Gaussian distribution, the parameters in the orthogonal matrix $x$ and $\gamma\equiv \cosh 2y$ are varied within the interval $0-2\pi$ and $0-50$ (as shown in the figures) with a flat distribution. In each of the cases, the magenta colour represents $N_{B-L}^{G0}$ whereas the red, blue and green are the representative colours for electron, muon and tau flavour asymmetries. It is evident that both the light neutrino mass ordering in case-a are ruled out.  In case-b, though the inverted ordering produces the $N_{B-L}^{G0}$ within the correct range, the individual flavour asymmetries, as well as the sum of the flavour asymmetries, unfortunately, struggle to reproduce the correct asymmetry. 

However, the normal mass ordering in case-b perfectly reproduces the observed asymmetry in all the flavours. An intriguing aspect is that the current neutrino oscillation data favour normal mass ordering and as we see, successful RIGL in minimalistic seesaw also requires normal mass ordering. We also note that the total asymmetry goes below the observed range for the value of $M_1\lesssim 6.3\times 10^6$ GeV.  In minimal seesaw, this can be regarded as the lower bound on $M_1$ for a successful RIGL. Note that now the bound is three orders of magnitude below the bound that is obtained in leptogenesis from RH neutrino decays ($M_{\rm  Lightest}\gtrsim 10^9\rm  GeV$).  Coming into a bit more detail regarding the plots, for the case-a, as explained previously that the dominant contribution in the flavoured asymmetries comes from the projection effect for which the flavoured asymmetries escape $N_2$ washout (cf. Eq.\ref{2fl}). Despite a single-stage washout, the flavoured asymmetries struggle to produce large baryon asymmetry due to lack of hierarchical enhancement. However, unlike case-a, the hierarchical enhancement in case-b is much more stronger, and one obtains correct baryon asymmetry. Regarding the nature of the plots, firstly, as shown in Fig.\ref{tri}, for normal light neutrino ordering the strength of the washout (inverse decay) in the electron flavour is much weaker than the other two flavours--making the RH neutrinos `electrophobic'. This is clearly visible in the upper panel. In both the plots, the parameter space, as well as the magnitude of the asymmetry, is dominated by electron flavour (red region). Also notice that the parameter spaces of muon and tau flavours are more or less similar since washout in these flavours are of equal strength (cf. Fig.\ref{tri}). However, for the inverted mass ordering the parameter space is dominated by the muon and tau flavour--a fact that was already anticipated in Fig.\ref{tri} (lower panel). Finally, the magnitude of the $N_{B-L}^{G0}$ as well the flavoured asymmetries get reduced for large values $\gamma$. This is simply due to the fact that for large values of $\gamma$, the parameter $\alpha$ increases (since $K_1$ increases) which causes a late freeze-out of $N_{B-L}^{G0}$ ($z_{\rm in}$ and $z_0$ are well separated\footnote{ This could be theoretically problematic for case-a since before the asymmetry freezes out $N_2$-washout starts to act and therefore Eq.\ref{n2w} is approximately valid. However, in any case, as we see, case-a lacks hierarchical enhancement and generates asymmetry several order of magnitude below $N_{B-L}^{\rm Obs}$. Thus this consideration of this type hardly matters.}) and hence a reduction in the magnitude of the final frozen out values of $N_{B-L}^{G0}$. For large values of $\gamma$ the magnitude of the flavoured decay parameters increases as well. This is the reason that the flavour asymmetries are more suppressed for large values of $\gamma$. This we think an interesting aspect of RIGL -- successful leptogenesis naturally requires {\it less fine-tuning in seesaw formula}. Note also that the condition of strong flavour effect is satisfied in the small-$\gamma$ region which is preferred by successful leptogenesis in RIGL. Before we conclude, we would like to make the following remarks (certainly not exhaustive): \\
$\bullet$ Should we wish to generalise this work into a three-RH neutrino scenario to probe a pure gravitational leptogenesis, we might consider two relevant hierarchical spectrum of masses I) $M_{2,3}\gg T_{\rm in}\equiv T_{\rm RH}$ and $M_1\ll 10^9$ GeV II) $M_{3}\gg T_{\rm in}\equiv T_{\rm RH}$ and $M_{1,2}\ll 10^9$ GeV. In the former case, we expect similar results as in the present scenario (now the hierarchical enhancement is controlled by $M_3/M_1$). However, in the  latter one we expect the lower bound on $M_1$ to be more stringent since the washout  would  be strong:  $e^{-(K_{1\alpha}+K_{2\alpha})}$ instead of $e^{-(K_{1\alpha}})$. However, to make a precise statement, this requires careful investigation.\\
$\bullet$ RIGL from low energy CP phases? As we see in Eq.\ref{nblg0} that $N_{B-L}^{G0}$ is free from the neutrino mixing matrix $U$ and thus in general RIGL is not directly connected to low energy CP phases. This is somewhat similar to nonthermal leptogenesis from inflaton decay\cite{rome}. However, there are well known techniques (models) to express the elements of $\Omega$ matrix in terms of low energy phases by reducing the number of parameters in seesaw models\cite{Frampton:2002qc,King:2013iva}. Keeping in mind significant progress in low energy neutrino experiments (neutrino parameters including the Dirac CP phase are getting measured with high statistical significance) models of these kinds are worth to explore in the light of RIGL which has never been done before. \\
$\bullet$ Imprints of RIGL on absolute neutrino mass scale and neutrino-less double beta decay? 
This point is a little bit tricky, and we have to opt for the paradigm of strong thermal leptogenesis\cite{Bertuzzo:2010et,DiBari:2014eqa} where we {\bf don't want} any pre-existing asymmetry to compete with the asymmetry that is produced by the decays. When $M_2$ is in the two flavour regime, and $M_1$ is in the three flavour regime, strong thermal leptogenesis can be successfully implemented with the following conditions:
\bea
K_{2\tau},K_{1e},K_{1\mu}\gg 1,~K_{1\tau}\ll 1.\label{st10}
\eea
By considering $K_{2\tau}\gg 1$, we erase $\tau$ component of $N_{B-L}^{G0}$ by washing it out. Note that $K_{2\mu,2e}\gg 1$ do not help. Since in this case though with these conditions we can washout the asymmetry in the direction of $\tau_2^\perp(e+\mu)$, the asymmetry orthogonal to  $\tau_2^\perp$ will survive and therefore one needs to wash it out at later stage by $N_1$ (washout effects act in all three directions of flavour) in the $e$ and $\mu$ flavours by choosing $K_{1e},K_{1\mu}\gg 1$. Therefore, successful leptogenesis can be done by the decays of $N_2$ (One now introduces $N_3$ as well to have sufficient CP violation) in the $\tau$ flavour, since we still have $K_{1\tau}\ll 1$--asymmetry generated by $N_2$ survives $N_1$-washout in the direction of $\tau$. Such a hierarchical mass splitting can naturally be generated in SO(10) models\cite{so5}, and the strong thermal conditions in Eq.\ref{st10}, in general, give lower bounds on $m_1$ and the neutrinoless double beta decay parameter $m_{ee}$ within the testable range of the cosmological and double beta decay experiments\cite{DiBari:2014eqa}.  However, in all the previous studies, the magnitude of the pre-existing asymmetries have been put by hand, i.e., the magnitude does not depend on seesaw model parameters. But now, in this case, the most interesting part is, $N_{B-L}^{G0}$ depends on the Yukawa couplings. Therefore, we expect the scenario would be more constrained than the previous studies. A dedicated analysis in this direction will be provided in a forthcoming publication\cite{last}.\\
$\bullet$ We have not discussed non-standard cosmological scenarios, e.g., a fast-expanding universe with an equation of state $1/3<\omega<1$ in which gravitational leptogenesis can be implemented even without hierarchical enhancement\cite{rg10}.\\

As a concluding remark, in seesaw models, RIGL mechanism opens up several interesting avenues which are worth exploring parallel to standard thermal leptogenesis from decays.

\section{Summary }\label{s5}
We discuss flavour effects on right-handed neutrino induced gravitational leptogenesis in the minimal seesaw model. We particularly consider two different spectrum of RH neutrino masses a)  $10^9 {\rm GeV}\ll M_2\ll 10^{12} {\rm GeV}\lesssim T_0$, $M_1\ll 10^9$  GeV, i.e., $M_2$ is in the two flavour regime and $M_1$ is in the three flavour regime  b) $ 10^{12} {\rm GeV}\lesssim T_0\ll M_2$, $M_1\ll 10^9$  GeV,.i.e., $M_2$ is in the unflavoured (one flavour) regime and $M_1$ is in the three flavour regime with $T_0$ being the temperature at which a frozen out lepton asymmetry is generated from gravitational mechanism.  For these spectrum of masses, observed baryon asymmetry cannot be generated by RH neutrino decays. We show that for the same spectrum of masses, unflavoured gravitational leptogenesis does not successfully reproduce the observed baryon asymmetry. However, when flavour effects on washout processes are taken into account, for the case-b, the observed baryon asymmetry could be generated by the gravitational mechanism. We also show that the lower bound on $M_1$, in this case is, $\mathcal{O}(10^6)$ GeV which is three orders of magnitude below than what is obtained from RH neutrino decays. We then discuss the future outlook of the gravitational leptogenesis mechanism, particularly its testability in low energy neutrino experiments.
\section*{Acknowledgement}
The authors would like to thank Graham M. Shore for helpful discussions and useful correspondence of Ref.\cite{rg10} while we were finishing this draft. RS is supported by  Newton International Fellowship (NIF 171202).
{}
\end{document}